%% file: main.tex
\documentclass[acmsmall]{acmart}
\AtBeginDocument{%
  \providecommand\BibTeX{{%
    \normalfont B\kern-0.5em{\scshape i\kern-0.25em b}\kern-0.8em\TeX}}}

\setcopyright{acmcopyright}
\copyrightyear{2018}
\acmYear{2018}
\acmDOI{XXXXXXX.XXXXXXX}
\acmJournal{JACM}
\acmVolume{37}
\acmNumber{4}
\acmArticle{111}
\acmMonth{8}

\usepackage[utf8]{inputenc}
\usepackage{adjustbox}
\usepackage{multicol}
\usepackage{booktabs}
\usepackage{diagbox}
\usepackage{multirow}
\usepackage{fancyvrb}
\VerbatimFootnotes
\usepackage{url}
\usepackage{listings}
\usepackage{subcaption}
\usepackage{xcolor}
\usepackage[most]{tcolorbox}
\usepackage{tabularx}
\usepackage[shortlabels]{enumitem}
\usepackage[flushleft]{threeparttable}

\makeatletter  
\newif\if@restonecol  
\makeatother

\usepackage[linesnumbered,ruled,vlined]{algorithm2e}
\SetKw{Continue}{continue}
\usepackage{algpseudocode}  
\usepackage{amsmath}  
\mathchardef\mhyphen="2D

\AtBeginDocument{%
  \providecommand\BibTeX{{%
    \normalfont B\kern-0.5em{\scshape i\kern-0.25em b}\kern-0.8em\TeX}}}

\setcopyright{acmcopyright}
\copyrightyear{2018}
\acmYear{2018}
\acmDOI{10.1145/1122445.1122456}

\begin{document}
\title{Duplicate Bug Report Detection: How Far Are We?}

\author{Ting Zhang}
\affiliation{%
  \institution{Singapore Management University}
  \country{Singapore}
}
\email{tingzhang.2019@phdcs.smu.edu.sg}

\author{DongGyun Han}
\affiliation{%
  \institution{Royal Holloway, University of London}
  \country{United Kingdom}
}
\email{DongGyun.Han@rhul.ac.uk}

\author{Venkatesh Vinayakarao}
\affiliation{%
    \institution{Chennai Mathematical Institute}
  \country{India}
}
\email{venkateshv@cmi.ac.in}

\author{Ivana Clairine Irsan}
\affiliation{%
    \institution{Singapore Management University}
  \country{Singapore}
}
\email{ivanairsan@smu.edu.sg}

\author{Bowen Xu}
\affiliation{%
    \institution{Singapore Management University}
  \country{Singapore}
}
\email{bowenxu@smu.edu.sg}
\authornote{Corresponding author.}

\author{Ferdian Thung}
\affiliation{%
    \institution{Singapore Management University}
  \country{Singapore}
}
\email{ferdianthung@smu.edu.sg}

\author{David Lo}
\affiliation{%
  \institution{Singapore Management University}
  \country{Singapore}
}
\email{davidlo@smu.edu.sg}

\author{Lingxiao Jiang}
\affiliation{%
\institution{Singapore Management University}
\country{Singapore}
}
\email{lxjiang@smu.edu.sg}

\renewcommand{\shortauthors}{Zhang et al.}

\begin{abstract}
Many Duplicate Bug Report Detection (DBRD) techniques have been proposed in the research literature.
The industry uses some other techniques. 
Unfortunately, there is insufficient comparison among them, and it is unclear how far we have been. 
This work fills this gap by comparing the aforementioned techniques. 
To compare them, we first need a benchmark that can estimate how a tool would perform if applied in a realistic setting today. 
Thus, we first investigated potential biases that affect the fair comparison of the accuracy of DBRD techniques. 
Our experiments suggest that data age and issue tracking system choice cause a significant difference. 
Based on these findings, we prepared a new benchmark. 
We then used it to evaluate DBRD techniques to estimate better how far we have been. 
Surprisingly, a simpler technique outperforms recently proposed sophisticated techniques on most projects in our benchmark. 
In addition, we compared the DBRD techniques proposed in research with those used in \texttt{Mozilla} and \texttt{VSCode}. 
Surprisingly, we observe that a simple technique already adopted in practice can achieve comparable results as a recently proposed research tool. 
Our study gives reflections on the current state of DBRD, and we share our insights to benefit future DBRD research.

\end{abstract}



\maketitle

\input{sec/1_introduction}
\input{sec/2_background}
\input{sec/3_approach}
\input{sec/4_experiment}
\input{sec/5_result}
\input{sec/6_discussion}
\input{sec/7_threats}
\input{sec/8_relatedwork}
\input{sec/9_conclusion}

\bibliographystyle{ACM-Reference-Format}
\bibliography{main}

\end{document}

%% file: sec/1_introduction.tex
\section{Introduction}
\label{sec:introduction}
Bug reports (BRs) play an essential role in the software development process, and meanwhile, issue tracking systems (ITSs) become necessary to manage BRs. 
The existence of duplicate BRs may cost extra software maintenance efforts in bug triage and fixing~\cite{kucuk2021characterizing}. 
In practice, duplicate BRs can also be hard to identify. 
For example, prior research ~\cite{rakha2016studying} found that there are duplicate BRs taking thousands of days to identify, with up to 230 comments, involving up to 75 people. 
This kind of BRs consumed much effort before it was finally identified as a duplicate.
To alleviate the heavy burden of triagers and decrease the cost of software maintenance, many automatic techniques have been proposed in the past decade~\cite{sun2010discriminative, zhou2012learning, alipour2013contextual, rodrigues2020soft}.
Among them, Duplicate Bug Report Detection (DBRD) techniques have been deployed in practice. 
For example, some ITSs provide recommendations of potential duplicates to a reporter prior to submitting a BR.
Others employ a bot that flags a duplicate after it has been submitted. 
For both scenarios, a ranked list of potential duplicate BRs is produced for manual inspection.

Despite the many research works and practitioners' adoption of DBRD, unfortunately, it is unclear which DBRD technique can recommend the duplicate BR most accurately overall. 
The most recent work by Rodrigues et al.~\cite{rodrigues2020soft} shows that \texttt{SABD}~\cite{rodrigues2020soft} outperforms \texttt{REP}~\cite{sun2011towards} and \texttt{Siamese Pair}~\cite{deshmukh2017towards}. 
However, their experiments are only limited to a collection of old BRs from Bugzilla ITSs, in which the latest data used belongs to the year 2008. 
Concurrent with the work by Rodrigues et al.~\cite{rodrigues2020soft}, Xiao et al.~\cite{xiao2020hindbr} and He et al.~\cite{he2020duplicate} have proposed other DBRD solutions. 
They have not been compared to each other. 
Besides, they did not compare with the tools used in practice.
This motivates us to: (1) create a benchmark that addresses the limitations of existing evaluation datasets, (2) compare research tools on the same dataset, and (3) compare research and industrial tools.

By studying evaluation data used in the literature, we identify three potential limitations:

\begin{itemize}[leftmargin=1em,nosep]
    \item{\textbf{Age bias:} Firstly, most of the techniques have not been evaluated on the recent BRs. Previous research relies heavily on the dataset proposed by Lazar et al.~\cite{lazar2014generating}. This dataset contains BRs from four projects (\texttt{Eclipse}, \texttt{Mozilla}, \texttt{Netbeans}, and \texttt{OpenOffice}) stored in their respective Bugzilla ITSs. All the BRs in these four projects are reported as early as July 1998 and till January 2014. The effectiveness of DBRD techniques on BRs submitted in the year 1998 should be significantly different from their effectiveness on recent BRs. Clearly, a DBRD technique that works well on data from 10 years ago but no longer so on recent data should not be of much use to developers today. We refer to this potential bias as {\em age bias}.}
    \item{\textbf{State bias:} Secondly, as indicated by Xia et al.~\cite{xia2014empirical}, several fields of a BR may change during its lifetime, e.g., \texttt{summary}, \texttt{version}, \texttt{priority}, etc. Various reasons can lead to changes in BR fields, such as when a bug reporter is new to the open-source project, he might submit a BR where some of the fields are wrongly assigned. Most studies evaluate the effectiveness of DBRD techniques based on the {\em latest} states of the fields at the point of data collection. The effectiveness of DBRD techniques should be significantly different if the {\em initial} states of the fields are used. We refer to this potential bias as {\em state bias}.}
    \item{\textbf{ITS bias:} Lastly, as these techniques are commonly evaluated on BRs from one specific ITS (i.e., Bugzilla), it is unclear how they perform on other ITSs (e.g., Jira and GitHub). These ITSs are different in the list of fields that they support, and some DBRD techniques make use of fields that exist in Bugzilla but not others. A DBRD technique that is specifically designed for a certain ITS data should perform differently when applied in another ITS. We refer to this potential bias as {\em ITS bias}.}
\end{itemize}

These biases motivate us to investigate our first research question:
\begin{tcolorbox}[left=0pt,right=0pt,top=0pt,bottom=0pt,boxrule=0pt, frame empty]
  \textbf{RQ1:} \textit{How significant are the potential biases on the evaluation of DBRD techniques?}
\end{tcolorbox}

To answer RQ1, we investigate the performance of the three best-performing solutions identified by Rodrigues et al.~\cite{rodrigues2020soft} in the presence and absence of each potential bias. We demonstrate that the {\em age bias} and {\em ITS bias} matter significantly ($p$-value<0.01) and substantially (large effect size) for all but one case, while the impact of {\em state bias} is insignificant ($p$-value>0.05) for all cases.

Based on the above findings, we create a benchmark that addresses age bias and ITS bias and use it to evaluate DBRD techniques that have been shown competitive performance in the research literature.
Specifically, we conduct a comparative study that evaluates DBRD techniques on recent BRs (addressing {\em age bias}) from different ITSs (addressing {\em ITS bias}).
Other than the three techniques mentioned before, we also include two additional recently-proposed DBRD techniques, \texttt{DC-CNN}~\cite{he2020duplicate}, and \texttt{HINDBR}~\cite{xiao2020hindbr}.
We hereby ask our second research question:
\begin{tcolorbox}[left=0pt, right=0pt, top=0pt, bottom=0pt, boxrule=0pt, frame empty]
  \textbf{RQ2:} \textit{How do state-of-the-art DBRD research tools perform on recent data from diverse ITSs?}
\end{tcolorbox}

Our result shows that, surprisingly, for most projects, the retrieval-based approach, i.e., \texttt{REP}~\cite{sun2011towards}, proposed a decade ago, can outperform the recently proposed more advanced and sophisticated models based on deep learning by $22.3\%$ on average in terms of Recall Rate at Top-10 positions (a.k.a.\ RR@$10$). 
This again demonstrates the value of simpler approaches in the saga of simple vs.\ complex~\cite{zeng2021deep,menzies2018500+,fu2017easy}.

Further, the research tools have been evaluated in isolation ignoring tools that have been used in practice.
To address this gap, we investigate our third research question: 

\begin{tcolorbox}[left=0pt, right=0pt, top=0pt, bottom=0pt, boxrule=0pt, frame empty]
  \textbf{RQ3:} \textit{How do the DBRD approaches proposed in research literature compare to those used in practice?}
\end{tcolorbox}

To answer RQ3, we compare the five research tools considered in RQ2 with two tools used by practitioners.
The first tool is the DBRD technique implemented in the Bugzilla ITS, named Full-Text Search (\texttt{FTS}).
It has been deployed in practice by \texttt{Mozilla}~\cite{bugzillamozilla}. 
The next tool is the \texttt{VSCodeBot}, which includes a DBRD feature used in Microsoft's \texttt{VSCode} repository~\cite{githubVSCode}.
The experimental results with \texttt{FTS} show that a straightforward duplicate search method used in \texttt{Mozilla} can act as a useful baseline as it can outperform the second best-performing technique on one of the projects in our benchmark by $7.6\%$ in terms of RR@$10$.
Still, the best-performing research tools can boost \texttt{FTS} performance by $22.1\%$ to $62.7\%$.
Moreover, the experimental results on \texttt{VSCodeBot} show that \texttt{VSCodeBot} is better than most tools.
Still, the two best-performing research tools can outperform \texttt{VSCodeBot} by $7.6\%$ and $9.8\%$ in terms of RR@$5$.
The results show the value of research on DBRD and the need to develop these research tools further so that practitioners can use and benefit from them.

Our main contributions can be summarized as follows:
\begin{itemize}[leftmargin=1em,nosep]
    \item \textit{A study about bias in DBRD data.} We investigate the significance of age bias, state bias, and ITS bias on the evaluation of DBRD techniques. The result depicts that age bias and ITS bias have significant and substantial impacts, while state bias has no significant impact.
    \item \textit{A new benchmark.} We provide a rich dataset containing recent three-year BRs from Bugzilla, Jira, and GitHub ITSs of six projects. We release our replication package including the data and code.~\footnote{\url{https://github.com/soarsmu/TOSEM-DBRD}}
    \item \textit{A comparative study.} We evaluate state-of-the-art DBRD research tools on a revised dataset that addresses age and ITS biases, and the result shows a simple retrieval-based approach can beat recently-proposed sophisticated deep learning-based models.
    We also compare state-of-the-art DBRD research tools with two industrial tools, i.e., (1) full-text search (\texttt{FTS}) implemented in Bugzilla ITS and used by \texttt{Mozilla}, (2) \texttt{VSCodeBot} used by \texttt{VSCode} repository. The result shows that FTS outperforms some research tools on some projects. The best-performing research tools however can outperform FTS by $22.1\%$ to $62.7\%$ in terms of RR@$10$. \texttt{VSCodeBot} is better than most research tools, but the best research tool can outperform it by $9.8\%$ in terms of RR@$5$.
\end{itemize}

The rest of the paper is organized as follows. 
Section~\ref{sec:background} presents the background information on DBRD. 
Section~\ref{sec:data} describes how we construct the dataset. 
Section~\ref{sec:experiments} details our experimental setup. 
Section~\ref{sec:results} reports and analyzes the experimental results. 
Section~\ref{sec:discussion} outlines the novel insights from our experiments. 
The analysis of threats to validity is covered in Section~\ref{sec:threats}. Section~\ref{sec:related} discusses the related work.
Finally, we conclude our work and outline the potential future research directions in Section~\ref{sec:conclusion}. 

%% file: sec/2_background.tex
\section{Background}
\label{sec:background}

Several DBRD techniques have been proposed in research to detect duplicate BRs automatically. 
Yet, the industry uses a different set of tools.
However, there is a lack of a systematic study about different approaches for DBRD.
Here, we present an overview of BRs in ITSs, DBRD in practice, and DBRD in research.

\subsection{Bug Reports in Issue Tracking Systems}
\label{sec:studing}

\begin{table}[t]
\begin{threeparttable}
\caption{Three example bug reports from \texttt{Eclipse} (Bugzilla), \texttt{Apache} (Jira), \texttt{VSCode} (GitHub) project.}
\label{tab:bug_examples} 
\centering
\small
\begin{tabularx}{\textwidth}[t]{|p{2cm}|X|X|X|}
    \hline
    \textbf{Field} & \textbf{Bugzilla} & \textbf{Jira} &  \textbf{GitHub} \\
    \hline
    \hline
    \textit{Bug Id} & 542516  & HIVE-21207 & 92171 \\
    \textit{Created} & 2018-12-07 08:01 EST & 04/Feb/19 11:02 & 7 Mar 2020  \\
    \textit{Product} & Platform & - & - \\ 
    \textit{Component} & SWT & None & - \\
    \textit{Version} & 4.8 & None & -\\
    \textit{Priority} & P3 & Major & -\\
    \textit{Severity} & major & - & - \\
    \textit{Status} & CLOSED  & RESOLVED & CLOSED \\
    \textit{Resolution} & DUPLICATE & Duplicate & - \\
    \textit{Summary} & Unable to uplaod Image after login & Use 0.12.0 libthrift version in Hive & "C\# extension recommended for this file type" \\
    \textit{Description} & Starting with Eclipse 4.8, horizontal scrolling is ...(\textit{omitted})
    &  Use 0.12.0 libthrift version in Hive. 
    & Issue Type: Bug Recently, when I open a .cs file. I get the notification... (\textit{omitted}) \\
    \hline
    \hline
    Ways to Record Duplicate & Field: <dup\_id>530693</dup\_id> & Issue Links: is duplicated by Bug HIVE-21173; HIVE-21000 & Infered from the comments below: \texttt{VSCodeBot} gave a recommendation, the issue reporter acknowledged\\
    \hline
    \hline
    \textit{URL:} & \url{https://bugs.eclipse.org/bugs/show_bug.cgi?ctype=xml&id=542516} & \url{https://issues.apache.org/jira/browse/HIVE-21207} & \url{https://github.com/microsoft/vscode/issues/92171}\\
    \hline
\end{tabularx}
\begin{tablenotes}
    \small
    \item `-' : this field is not available in the ITS; `None' : the value of the corresponding field is empty.
\end{tablenotes}
\end{threeparttable}
\end{table}

Existing research~\cite{rodrigues2020soft,xiao2020hindbr} heavily uses data from Bugzilla for comparing DBRD techniques. 
We observe differences across ITSs in submitting BRs, finding duplicates, and marking BRs as duplicates.
We select the ITSs in the study based on two factors: whether open-source projects can use them and their popularity. 
Based on the criteria, we choose Bugzilla, Jira, and GitHub.

\textbf{Bugzilla} is one of the world-leading free ITSs. 
Several large projects such as \texttt{Mozilla} and \texttt{Eclipse} use Bugzilla. 
A typical bug reporting process involves providing necessary textual and categorical details. 
A new BR submission in Bugzilla can be seen as a form-filling activity. 
To mark a BR as duplicate~\cite{bugzillafaq} in Bugzilla, users set the \texttt{resolution} field to \texttt{duplicate} and insert the bug id that this BR duplicates.

\textbf{Jira} has gained popularity over the last decade. 
According to the Jira official website~\cite{jirahome}, more than 65,000 organizations used Jira in 2021. 
Jira follows a form-filling design similar to Bugzilla for creating a new BR. 
While creating a BR, Jira allows users to relate existing bugs using labels: \texttt{is caused by}, \texttt{is duplicated by}, or \texttt{relates to}. 
New duplicates are manually marked using the \texttt{resolution} field.

\textbf{GitHub} is a popular Git repository hosting service. 
It provides rich features, one of which is issue tracking. 
Issues in GitHub carry a simpler structure when compared to Bugzilla and Jira. 
Apart from the textual information, the categorical fields are customizable for each repository. 
Repositories on GitHub can use \texttt{label} to categorize issues. 
However, these labels are not mandatory or well-defined like Bugzilla and Jira.
Hence, BRs logged in Bugzilla and Jira have relatively more categorical or structured information. 
This type of flexibility has pros and cons. 
On the one hand, it makes it easy to report the issue details; on the other hand, it makes it difficult to extract useful categorical information systematically.
Some GitHub repositories provide templates to help users embed categorical information in the textual information.
For example, the \texttt{VSCode} project provides a textual template that asks users to provide categorical information such as {\it VSCode version} and {\it os version}.
However, the textual template is not compulsory to follow and can be ignored by issue reporters.
The ITS of GitHub uses labels and comments to mark issues as duplicates.
We will describe its duplication marking approach in detail in Section~\ref{subsec:duplicate}.

BRs from different ITSs carry some common fields (such as \textit{BugId}, \textit{Created Date}, etc.) and also some different ones (such as \textit{Severity} that exists in BRs from Bugzilla ITS but does not exist in BRs from Jira and GitHub ITSs).
We show one example BR from each of the three ITSs in Table~\ref{tab:bug_examples}.
As shown in the table, a Bugzilla BR consists of several fields, the types of which can either be textual or categorical.
For categorical fields: \texttt{Product} usually represents a software product that is shipped, and \texttt{Component} is a part of a product.
\texttt{Severity} states how severe the problem is, while \texttt{Priority} is used by the bug assignee to prioritize the issues. 
\texttt{Status} indicates the current state the bug is in, and \texttt{Resolution} indicates what happens to this bug. 
A BR can have several possible \texttt{statuses}, such as \texttt{unconfirmed}, \texttt{confirmed}, \texttt{fixed}, \texttt{in process}, \texttt{resolved}, etc.
The possible \texttt{resolution} values of a bug can be \texttt{duplicate}, \texttt{fixed}, \texttt{wontfix}, etc.
However, Jira does not provide exactly the same fields. 
There are no \texttt{Severity} and \texttt{Product} fields in Jira. 
For GitHub, there are no categorical fields. The textual data mainly includes a \texttt{summary}, \texttt{description}, etc. 
All the ITSs contain these two common textual fields. The \texttt{Summary} is usually a one-sentence text describing a bug. 
The \texttt{Description} usually contains more details, such as the steps to reproduce the bug.

\subsection{DBRD in Practice}
\label{sec:dbrdinpractice}

\begin{figure}[t]
\includegraphics[width=0.8\linewidth]{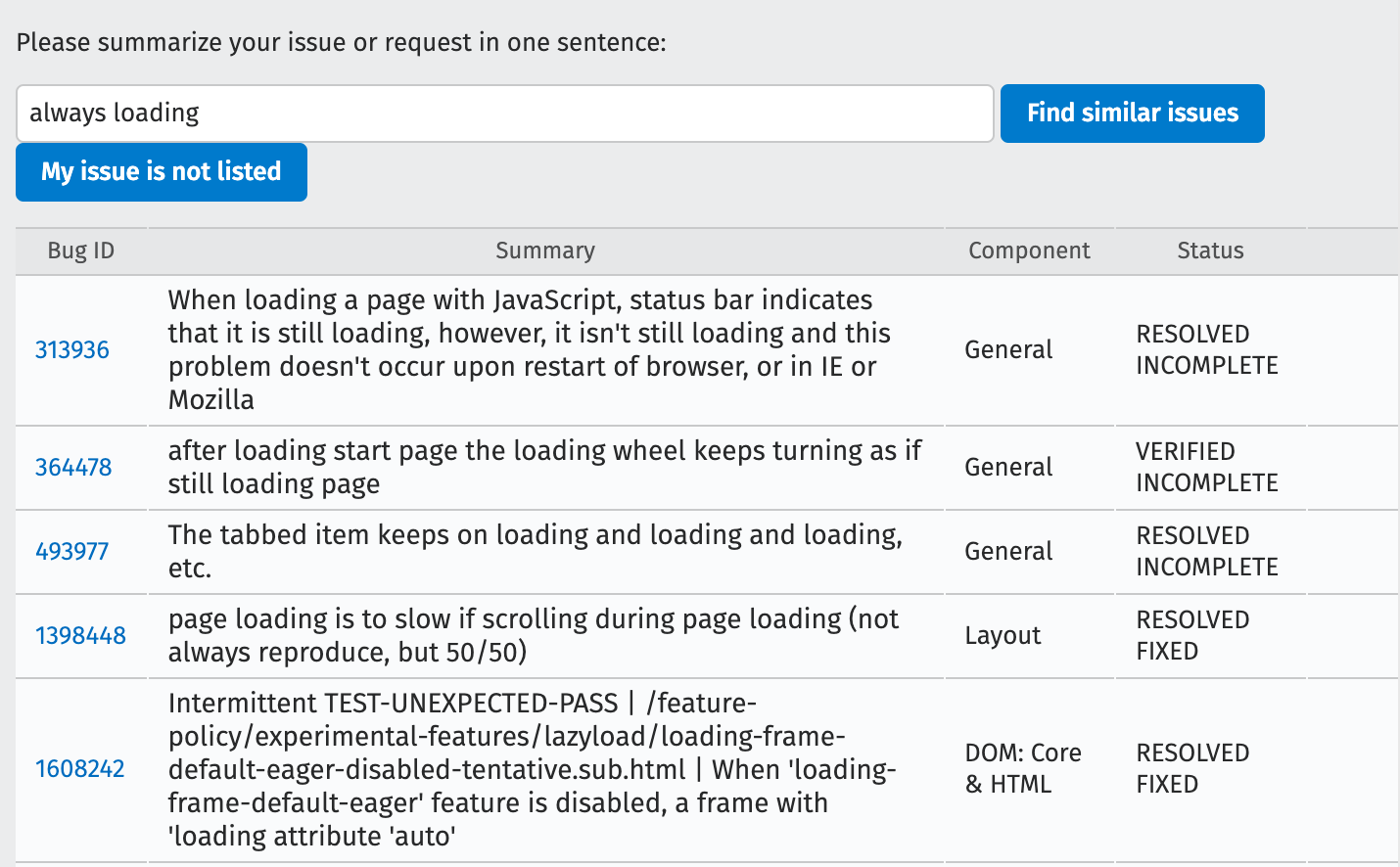}
\caption{An example of duplicate issue recommendation when typing ``always loading'' before submitting a new bug report on Mozilla Firefox}
\label{fig:mozilla}
\end{figure}

\begin{figure}[t]
\includegraphics[width=0.8\linewidth]{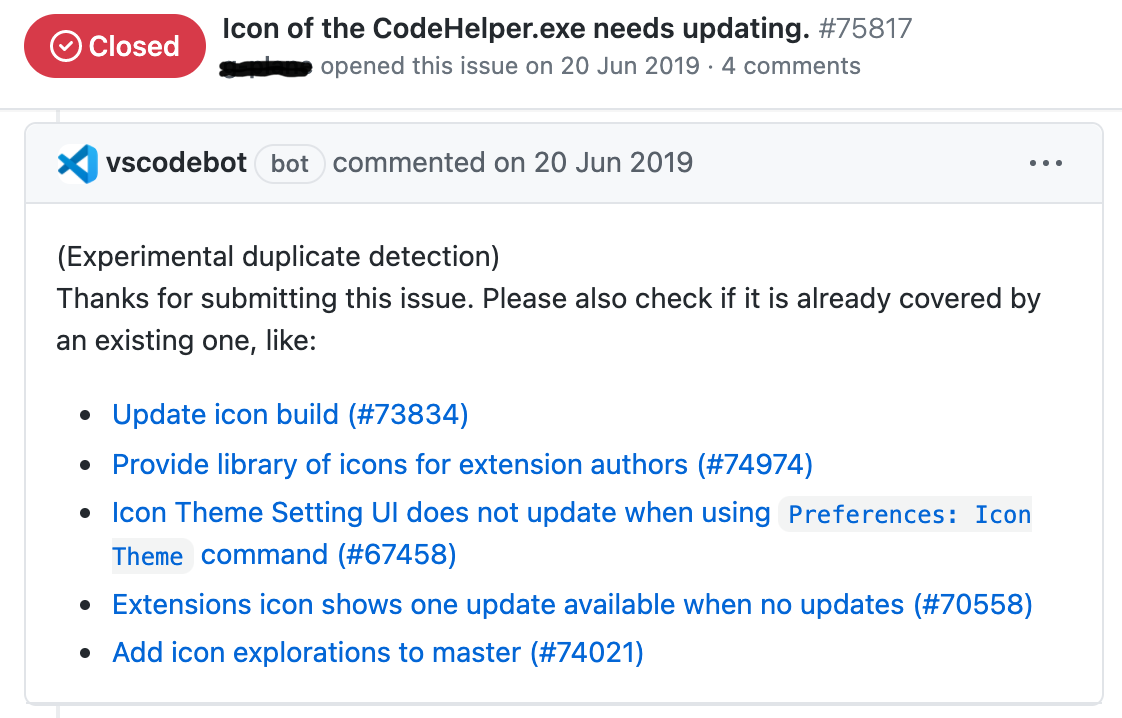}
\caption{An example of \texttt{VSCodeBot} duplicate issue recommendation (issue 75817) from Microsoft/VSCode repository}
\label{fig:VSCodebot}
\end{figure}

The existence of duplicate BRs causes increasing software maintenance efforts in bug triage and fixing~\cite{kucuk2021characterizing}.
Several causes can result in duplicate BRs. 
Prior work~\cite{bettenburg2008duplicate} shows that duplicate BRs can either be submitted intentionally (e.g., usually when the reporters are frustrated by the same bug not being resolved), or unintentionally (e.g., reporters do not search for existing BRs or cannot identify duplicates due to the lack of experience). To help lighten the workload of bug triagers and avoid redundant bug fixing, DBRD techniques can be useful in two use scenarios. DBRD approaches can either work for (1) \textit{pre-submission} scenario: They can be integrated into ITSs to make bug reporters aware of existing similar BRs and prevent duplicate BRs from happening. Figure~\ref{fig:mozilla} depicts how the JIT duplicate recommendation works at the pre-submission for \texttt{Mozilla} ITSs. or (2) \textit{post-submission} scenario: They can recommend duplicate lists after duplicate BRs are submitted.
For example, \texttt{VSCodeBot} that is adopted by \texttt{VSCode} GitHub repository can recommend a list of potential duplicates.
Like other bots~\cite{golzadeh2021identifying}, \texttt{VSCodeBot} also communicates with developers via issue comments and pull request comments. Figure~\ref{fig:VSCodebot} shows an example of \texttt{VSCodeBot} duplicate issue recommendation on issue $75817$\footnote{\url{https://github.com/microsoft/vscode/issues/75817}}. 
After a user submits the issue, \texttt{VSCodeBot} detects five potential duplicates. 
The two usage scenarios may require different technical considerations. 
For the pre-submission usage scenario, the efficiency of DBRD tools plays an important role as it requires DBRD tools to produce a real-time recommendation.
Issue reporters are unlikely to be willing to wait for a long time for a DBRD tool to return some results. 
On the other hand, for the post-submission usage scenario, the DBRD tools can potentially be run overnight in a batch mode to process many BRs, and the emphasis is on optimizing accuracy.

\subsection{DBRD in Research}
\label{sec:dbrdinresearch}
Here, we present an overview of the DBRD research tools considered in our work.
In the research literature, many DBRD approaches have been proposed and evaluated. 
In our work, we consider five approaches. 
They include the three best performers in the study conducted by Rodrigues et al.~\cite{rodrigues2020soft}: \texttt{SABD}~\cite{rodrigues2020soft}, \texttt{REP}~\cite{sun2011towards}, and \texttt{Siamese Pair}~\cite{deshmukh2017towards}.
\texttt{REP} is a popular information retrieval-based DBRD approach. 
\texttt{Siamese Pair} is the first deep learning-based DBRD approach proposed in the literature. 
We further include \texttt{HINDBR}~\cite{xiao2020hindbr} and \texttt{DC-CNN}~\cite{he2020duplicate}, which are proposed recently. 
Since DBRD models mainly differ in how they conduct feature engineering and how they measure similarity between BR pairs, we distinguish these two aspects between different methods in Table~\ref{tab:model_comp}.
For detailed explanations of each model, we refer readers to check the original papers. 
We describe the five approaches in the following paragraphs respectively.

\begin{table}[t]
\footnotesize
\caption{Comparison between different approaches} 
\label{tab:model_comp} 
\centering
\begin{tabular}{|l|c|c|c|c|}
    \hline
    \multirow{2}{*}{\textbf{Approach}} &  
    \multirow{2}{*}{\textbf{Type}} &
    \multicolumn{2}{c|}{\textbf{Feature Engineering}}  &  
    \multirow{2}{*}{\textbf{Distance Measurement}} \\
    \cline{3-4} & & \textbf{Embedding} & \textbf{Modeling} &\\
    \hline
    \hline
    \multirow{2}{*}{\texttt{REP}~\cite{sun2011towards}} & 
    Categorical & - & handcrafted &   \multirow{2}{*}{linear combination} \\
    \cline{2-4}
    & Textual & - & $BM25F_{ext}$ &   \\
    \hline
    \texttt{Siamese}
    & Categorical &customized & single-layer & \multirow{2}{*}{Cosine Similarity} \\
    \cline{2-4}
    \texttt{Pair}~\cite{deshmukh2017towards}& Textual & GloVe   & bi-LSTM + CNN  &  \\
    \hline
    \multirow{2}{*}{\texttt{SABD}~\cite{rodrigues2020soft}} & Categorical & customized & \texttt{ReLU} & fully-connected layer\\
    \cline{2-5}
    & Textual & GloVe  & bi-LSTM + attention & fully-connected layer  \\
    \hline
    \multirow{2}{*}{\texttt{HINDBR}~\cite{xiao2020hindbr}} & Categorical & HIN2vec & MLP & \multirow{2}{*}{Manhattan Distance}  \\
    \cline{2-4}
    & Textual & Word2vec & RNN &   \\
    \hline
    \multirow{2}{*}{\texttt{DC-CNN}~\cite{he2020duplicate}} & Categorical & \multirow{2}{*}{Word2vec} & \multirow{2}{*}{dual-channel CNN} & \multirow{2}{*}{Cosine Similarity} \\
    \cline{2-2}
    & Textual & &  & \\
    \hline
\end{tabular}
\end{table}

\texttt{REP}~\cite{sun2011towards} is a retrieval function to rank BRs based on their similarity with an incoming BR.
\texttt{REP} considers 7 features, including 2 textual features, i.e., summary and description fields, and 5 categorical features, i.e., product, component, type, priority, and versions.
The similarity between the two BRs is a weighted linear combination of the scores of the 7 features. 
Specifically, \texttt{REP} represents texts with uni-grams and bi-grams and calculates the textual similarity by computing an extended version of BM25F~\cite{robertson2004simple}.
Among the five categorical features, if the values of the product, component and type fields of the two BRs are the same, the corresponding feature is 1; otherwise, the feature value is 0. 
For the remaining two features, i.e., priority and version, the corresponding feature is represented by the inverse of the distance between the values of priority and version of the two BRs. 
REP has a total of 19 parameters to be tuned, such as the weights of features used.
REP proposes to learn these parameters for the bug repositories under consideration using stochastic gradient descent by analyzing a training dataset of historical BRs. 
The training set of \texttt{REP} is a set of triples $(q, rel, irr)$, where $q$ is the query bug, $rel$ is a duplicate bug with $q$, and $irr$ is a non-duplicate bug with $q$.

\texttt{Siamese Pair}~\cite{deshmukh2017towards} is a deep learning architecture combining Siamese LSTM and CNN for DBRD.
\texttt{Siamese Pair} first encodes different types of features separately.
Specifically, for textual fields, the summary is encoded by a bidirectional LSTM, and the description is encoded by a CNN. 
The categorical fields, such as product, priority, and component, are encoded by a feed-forward neural network. 
In the end, the outputs from the three encoders would be concatenated to represent a BR. 
The proposed model uses Siamese neural networks and is trained with a max-margin objective. 
In the evaluation stage, BRs in the test dataset are sorted based on the Cosine similarity with the encoding of the query BR.

Soft Alignment Model for Bug Deduplication (\texttt{SABD})~\cite{rodrigues2020soft} receives a pair of BRs, a query BR, and a candidate BR. 
\texttt{SABD} is composed of two sub-networks: one for categorical information and the other for textual information. 
Each sub-network represents one type of information separately and uses a comparison layer to produce a comparative representation of the two BR vectors either in terms of the values of their categorical or textual fields, respectively. 
The categorical sub-network is a straightforward dense neural network, while the textual sub-network adopts a sophisticated architecture, where the core is the soft alignment comparison layer. 
The outputs of the two sub-networks are concatenated. 
A classifier layer receives the concatenated output and produces the final predicted probability regarding whether the candidate BR is the duplicate of the query BR. The soft-attention alignment is used to exchange the information between BRs before encoding the textual fields into a fixed-size vector.

\texttt{HINDBR}~\cite{xiao2020hindbr} represents BRs as a Heterogeneous Information Network (HIN).
The BRs are connected through their categorical fields.
For example, two BRs with bug id \textit{x} and \textit{y} that have the same priority, say \textit{Pri\_High}, are connected as \textit{x} -- \textit{Pri\_High} -- \textit{y} in the HIN.
Manhattan distance on the \texttt{HIN2Vec}~\cite{fu2017hin2vec} embedding is used to find semantic similarity in BRs.
Xiao et al. introduced three variations of \texttt{HINDBR}: (1) only use unstructured features (i.e., text), (2) only use structured features (i.e., categorical information), and (3) use both.  
In this study, we found that the models only use unstructured features show better performance than both, thus, we adopted the variant with only text.

Dual-Channel Convolutional Neural Networks (\texttt{DC-CNN})~\cite{he2020duplicate}  represents a BR pair with dual-channel matrices.  
\texttt{DC-CNN} extracts the values of four fields in a BR, i.e., product, component, summary, and description, and treats them as text. After the pre-processing, including tokenization, stemming, and stop word removal, a word2vec ~\cite{lilleberg2015support,rong2014word2vec} model is learned to capture semantic information in the BRs. \texttt{DC-CNN} converts each BR from text to a single-channel matrix based on the word representations learned by the word2vec model. To represent a BR pair, it combines two single-channel BR representation matrices into dual-channel matrices. The BR pair representations are then fed into a CNN model to capture the correlated semantic relationships between BR pairs. The output of the last layer of the model is used as the predicted similarity score between two BRs.

\texttt{REP}, \texttt{SABD}, and \texttt{Siamese Pair} were evaluated using a {\em ranking} setting: given a new BR, rank the potential duplicate BRs. 
\texttt{HINDBR} and \texttt{DC-CNN} were evaluated using a {\em classification} setting: given a pair of BRs, decide whether they are duplicates. 
Rodrigues et al.~\cite{rodrigues2020soft} claim that this classification setting is quite unrealistic since the real scenario presents a much larger set of negative candidates. 
Furthermore, when a new BR is submitted, all previously submitted reports are duplicate candidates. 
Thus, they believe the classification-setting highly overestimated performance. Following them and considering how \texttt{DBRD} is used in practice (c.f. Section~\ref{sec:dbrdinpractice}), we evaluate all these approaches using the {\em ranking} setting. 

%% file: sec/3_approach.tex
\section{Data Collection Methodology}
\label{sec:data}

Different ITSs represent BRs in different ways, and it is essential to prepare a unified data format to represent the BRs extracted from all of these ITSs. 
We present our methodology to build the BR dataset from three different ITSs. 
The methodology involves five main steps: (1) crawling BRs from the website, (2) filtering out open/unresolved BRs, (3) extracting duplicate BR relations, (4) extracting both textual and categorical information, and (5) cleaning and generating duplicate pairs. We elaborate on these steps in the following subsections.

\subsection{Data Source Selection and Crawling}

\paragraph{Project Selection.} From each of the three ITSs, we have selected two projects. Here, we describe the projects selected and the rationale behind the selection.

\vspace{0.2cm}\noindent{\bf Bugzilla}: We choose \texttt{Eclipse}~\cite{bugzillaeclipse} and \texttt{Mozilla}~\cite{bugzillamozilla} for Bugzilla, which have the two largest number of issues in the dataset of Lazar et al.~\cite{lazar2014generating}. \texttt{Eclipse} is an Integrated Development Environment (IDE), and its Bugzilla contains several products, including C Development Tools, Eclipse Modeling Framework, and so on. 
Similarly, the Mozilla software foundation also includes several projects, including the popular Firefox web browser. 

\vspace{0.2cm}\noindent{\bf Jira}: Following Xie et al.~\cite{xie2018detecting}, we select two Apache projects hosted on Jira\footnote{Although Jira is a proprietary bug tracking and project management software, it is free for open-source projects (e.g., the Apache Software Foundation projects) that meet certain criteria~\cite{atlassian}.}, i.e., \texttt{Hadoop}~\cite{jirahadoop} and \texttt{Spark}~\cite{jiraspark}. \texttt{Hadoop} provides a big data framework for distributed data storage and processing.
There are several projects that have been categorized as \texttt{Hadoop} in Jira.\footnote{\url{https://issues.apache.org/jira/secure/BrowseProjects.jspa?selectedCategory=10292}}
In this work, we include \texttt{Hadoop Common}, \texttt{Hadoop HDFS}, \texttt{Hadoop Map\/Reduce}, \texttt{Hadoop YARN}, \texttt{HBase}, \texttt{Hive}, and \texttt{Hadoop Development Tools} projects.
We collectively call them as \texttt{Hadoop}. 
\texttt{Spark} is an analytics engine for large-scale data processing.

\vspace{0.2cm}\noindent{\bf GitHub}: From the datasets listed at GHtorrent~\cite{Gousi13}, we choose the repositories having the largest number of issues. 
We use the latest update of this \texttt{GHtorrent} dataset dated March 6, 2021. 
Then we manually confirm whether each repository is still in active use. 
After excluding test or unavailable repositories, we got the five repositories that contained the largest number of issues. 
They are: \texttt{nixos/nixpkgs}, \texttt{microsoft/vscode}, \texttt{elastic/kibana}, \texttt{kubernetes/kubernetes}, and \texttt{ansible/ansible}.
Among these five repositories, \texttt{VSCode}~\cite{githubVSCode} and \texttt{Kibana}~\cite{githubkibana} contained the largest number of duplicate issues in 2018 -- 2020, so we included these two projects in our dataset.
Visual Studio Code (\texttt{VSCode}) is a popular multi-platform source-code editor provided by Microsoft. 
\texttt{Kibana} is a proprietary data visualization dashboard software for Elasticsearch, which is a search engine based on Lucene.

\paragraph{Time Range Selection.} We select two different time periods (i.e., old and recent) to investigate whether the age of data impacts the performance of DBRD techniques. For \textit{old} data, we choose January 1, 2012 to December 31, 2014. For \textit{recent} data, we include the recent three-year data from January 1, 2018 to December 31, 2020. The DBRD feature described in Section~\ref{sec:dbrdinpractice} was introduced to Bugzilla in 2011. Moreover, the feature adoption date for the projects is unclear. As we would like to focus on analyzing the impact of {\em age bias} (rather than the impact of DBRD feature introduction), we intentionally picked the time range after this feature was introduced to Bugzilla. Intuitively, if a significant age bias exists considering a 6-year gap period, the bias would have been more pronounced if a longer gap period is considered.

\paragraph{Crawling.} 
For both Bugzilla and Jira, we used the XML export API of the ITSs.
For GitHub issues, we used GraphQL API~\cite{githubgraphql} for retrieving the issues in JSON format.
Both XML export API and GraphQL API are publicly available.
We crawl the issues in June 2021, which is six months later than the aforementioned recent data, to minimize the number of open or unresolved issues.

\subsection{Filtering Out Open/Unresolved BRs}
Following Lazar et al.~\cite{lazar2014generating}, we only keep \textit{closed} or \textit{resolved} BRs among all the crawled BRs.
A typical life cycle of a bug can be abstracted into six main steps~\cite{weiss2007long}: \texttt{unconfirmed}, \texttt{new}, \texttt{in progress}, \texttt{resolved}, \texttt{verified}, and \texttt{closed}.
Note that a \textit{resolved} or \textit{closed} bug can be reopened in the future. Even so, the detailed life cycle in different ITSs may be different. 
Based on the definition of open bugs by Mozilla~\footnote{\url{https://wiki.mozilla.org/BMO/UserGuide/BugStatuses}}, we consider a BR as closed if (1) its status is either \texttt{resolved} or \texttt{verified}; or (2) its resolution is one of the seven types: \texttt{fixed}, \texttt{invalid}, \texttt{wontfix}, \texttt{moved}, \texttt{duplicate}, \texttt{worksforme}, and \texttt{incomplete}. For Eclipse, we add another possible status \texttt{CLOSED} as closed bugs based on the Eclipse Wiki~\footnote{\url{https://wiki.eclipse.org/Bug_Reporting_FAQ#What_is_the_life_cycle_of_a_bug_report.3F}}. 
For Jira projects, a closed bug has much more possible resolution types than Bugzilla. 
Thus, we regard a BR as closed if its \textit{status} is marked as either \texttt{resolved} or \texttt{closed}. 
For GitHub projects, as it only has \textit{state} field, other than resolution and status. 
We consider an issue as closed, if its state is \texttt{closed}.

\subsection{Identifying Duplicate BRs}
\label{subsec:duplicate}
The process to identify whether a BR is a duplicate one is referred to as {\em duplicate detection}. We use the following strategies to extract ground truths:

\vspace{0.2cm}\noindent{\bf Bugzilla:} The XML format of each issue in Bugzilla has a field called \texttt{dup\_id}, which contains a reference to a bug that the current bug is a duplicate of. The dup\_id can either be empty or contains another bug's bug\_id. If it is not empty, the current bug is a duplicate of the bug in dup\_id.
Therefore, we directly extract the \textit{dup\_id} of those bugs that have a resolution of \texttt{DUPLICATE} to gather the duplicate relation.

\vspace{0.2cm}\noindent{\bf Jira:} Jira does not provide the \texttt{dup\_id} like Bugzilla does. However, Jira provides rich types of issue dependency links, e.g., \texttt{duplicates}, \texttt{is duplicated by}, \texttt{contains}, to represent the relationship between issues. 
We identify the duplicate relations by analyzing the \texttt{duplicates}, and \texttt{is duplicated by} links.

\vspace{0.2cm}\noindent{\bf GitHub:} Other than the prior two ITSs, the issues in GitHub have a more flexible format. GitHub supports marking duplicates with a comment \cite{githubduplicate} of the format ``Duplicate of \#ISSUE\_NUMBER''. Such comment format can be used to mark the issue that is a duplicate of another issue. As it is not mandatory, we found that not all issue reporters strictly follow this comment format pattern, some users use short-hand notations to mark duplicates, for example, \textit{dup with \#ISSUE\_NUMBER}. In the end, we use a regular expression that can check the variations of duplicate comments to detect duplicate issues.
To validate the reliability of our regular expressions, we manually investigated the duplicate issues detected by the regular expression.
We randomly sampled 384 duplicate issue pairs based on the regular expressions from \texttt{Kibana} and \texttt{VSCode}, respectively.
Then, two authors (i.e., investigators) evaluated independently whether the extracted duplicate issues are real duplicates.
If there is a disagreement between the investigators, they discuss their investigation results until they reach an agreement.
We found only 3 and 21 cases were wrong (i.e., the extracted issues based on the regular expression are not real duplicates) for \texttt{Kibana} and \texttt{VSCode}, respectively.
Thus, our regular expression can detect duplicate issues with at least 94.5\% of accuracy.
Note we do not check false negatives. 
As for projects which use Bugzilla and Jira as ITS, false negatives are also possible since people may not mark the duplicates as such.

\subsection{Information Extraction}

\begin{table}[t]
\footnotesize
\caption {Textual and categorical fields that are leveraged by the approaches} 
\label{tab:categorical_info} 
\centering
\begin{tabular}{|l|l|c|c|c|c|c|}
    \hline
    \multicolumn{2}{|c|}{Fields} & 
    \textbf{\texttt{REP}}~\cite{sun2011towards} & \textbf{\texttt{Siamese-Pair}}~\cite{deshmukh2017towards} & \textbf{\texttt{SABD}}~\cite{rodrigues2020soft} & \textbf{\texttt{HINDBR}}~\cite{xiao2020hindbr} & \textbf{\texttt{DC-CNN}}~\cite{he2020duplicate}  \\
    \hline
    \hline
    \multirow{2}{*}{Textual} & summary & \checkmark & \checkmark & \checkmark & \checkmark & \checkmark\\
    \cline{2-7}
    & description & \checkmark & \checkmark & \checkmark & \checkmark & \checkmark \\
    \hline
    \multirow{6}{*}{Categorical} & product & \checkmark &\checkmark & \checkmark & \checkmark & \checkmark\\
\cline{2-7}
& component & \checkmark &\checkmark & \checkmark & \checkmark & \checkmark \\
\cline{2-7}
& priority & \checkmark &\checkmark & \checkmark & \checkmark & \\
\cline{2-7}
& severity & &\checkmark & \checkmark & \checkmark & \\
\cline{2-7}
& type & \checkmark & & & & \\
\cline{2-7}
& version & \checkmark & & & \checkmark & \\
\hline
\end{tabular}
\end{table}

All the selected DBRD approaches have taken advantage of both textual and categorical information to improve their effectiveness.
Table~\ref{tab:categorical_info} shows the fields that each approach can utilize.
The three ITSs have different fields to record categorical information. We extract all the essential fields needed by the approaches, i.e., bug id, categorical fields: \texttt{product}, \texttt{component}, \texttt{severity}, \texttt{priority}, \texttt{version}, \texttt{status}, \texttt{resolution}; textual information: \texttt{summary}, \texttt{description}. We describe how we extract such information from different ITSs.

\vspace{0.2cm}\noindent{\textbf{Bugzilla}}: A BR on Bugzilla that is exported as an XML file contains clear and well-defined field names. Around 20 fields are given in the XML file, however, not all of them are needed for DBRD. 
We thus parsed the XML file and saved the essential fields.

\vspace{0.2cm}\noindent{\textbf{Jira}}: Similar to Bugzilla, Jira has several pre-defined fields for bug reporters to fill in, e.g., \texttt{component}, \texttt{priority}, \texttt{version}. However, Bugzilla and Jira do not have identical categories. For instance, Jira only has a \texttt{priority} field to indicate the importance of an issue, while Bugzilla has \texttt{severity} and \texttt{priority}. Different from Bugzilla, BRs on Jira do not have the \texttt{product} and \texttt{severity} fields. If any field value is missing, we leave it as an empty string.

\vspace{0.2cm}\noindent{\textbf{GitHub}}:
As mentioned before, GitHub does not provide any well-defined fields for categorical information.
Even though labels may provide categorical information, labels are highly customizable for each project.
In addition, labels are shared by the other features (e.g., pull requests and discussion) of GitHub\footnote{\url{https://docs.github.com/en/issues/using-labels-and-milestones-to-track-work/managing-labels}}, so it does not represent categories for issues only.
Due to the above limitations, we only extracted textual information from GitHub issues.
We leave it as future work to investigate how to derive categorical information from GitHub issues.

\subsection{Data Cleaning}
\textit{Handling duplicates.} It is not rare that one BR has more than one duplicate BRs. For example, the bug 92250 has 45 duplicates\footnote{\url{https://bugs.eclipse.org/bugs/duplicates.cgi}}. A  \textit{group} or \textit{bucket} refers to a set of BRs which are duplicates to each other. The \textit{master} BR is the one to which the rest reports refer. Similar to the prior work~\cite{sun2011towards, rodrigues2020soft}, we also make the first submitted BR as the \textit{master} and the rest in the bucket as \textit{duplicates}. As we do not cover all the BRs in each ITS and only use the time range of three years, the master BRs of some duplicate BRs can be out of our considered time range (i.e., before January 1, 2012, or January 1, 2018). In that case, we choose the oldest BR in the bucket within the time range as a new master. By this cleaning step, the number of duplicates in our dataset has been smaller than the number of BRs that have been resolved as duplicates.

\textit{Basic pre-processing.} We conducted data pre-processing in the textual part (summary and description) of a BR: (1) removing punctuations, (2) lower-casing, (3) removing numbers, (4) removing stop words, (5) stemming, and (6) removing single characters.
We then tokenized the processed text with white spaces.
We reused the script in the replication package provided by Rodrigues et al.~\cite{rodrigues2020soft}.

\textit{Train-test split.} 
We first sort all the BRs chronologically.
Then, we select three years of data.
The first two years of data is used for training (including validation, if applicable), while the last year of data is used for testing. Except REP, all the other approaches need validation data. \texttt{SABD} and \texttt{Siamese Pair} use the last 5\% data in training data as validation data, while \texttt{HINDBR} and \texttt{DC-CNN} use 20\% and 10\% of the training pairs as validation pairs, respectively.

\textit{One-year time window.}
For each duplicate BR in the test data, its candidate duplicates are the BRs submitted within one year. We call it \textit{one-year time window} search range. Rodrigues et al. found that despite some minor differences, the findings using a one-year time window are similar to the ones with a longer frame of three years. Thus, we decide to use a one-year time window in our experiments.

\textit{Pairs generation.}
Generating duplicate and non-duplicate pairs is an important implementation detail integrated into each approach. 
By design, different approaches utilize different ratios of non-duplicate and duplicate pairs. 
And we follow the original implementation of each approach. Specifically, \texttt{REP} generates 30 times more non-duplicate pairs than duplicate pairs. 
\texttt{Siamese Pair} and \texttt{SABD} generate the same number of non-duplicate pairs as duplicate pairs. 
\texttt{HINDBR} and \texttt{DC-CNN} generate four times more non-duplicate pairs than duplicate pairs. However, the duplicate pairs in the training data are the same, i.e., pair two BRs that are in the same bucket. 
We report the number of duplicate pairs in training data in Table~\ref{tab:data_six_projs}. 
Note that we would only pair those BRs in the training set.

\begin{table*}[t]
\begin{threeparttable}
\caption {Statistics of data in the six projects}
\label{tab:data_six_projs} 
\centering
\small
\begin{tabular}{|l|l|l|l|l|l|l|l|l|l|}
    \hline
    \multirow{2}{*}{\textbf{ITS}} & \multirow{2}{*}{\textbf{Project}} & \multirow{2}{*}{\textbf{\# BRs}} & \multirow{2}{*}{\textbf{\# Dup BRs (\%)}} & \textbf{\# Unique} & \multicolumn{2}{c|}{\textbf{Per bucket}} \\
    \cline{6-7}
    & & & & \textbf{Master BRs} & \textbf{Avg. BRs} & \textbf{Max. BRs}\\
    \hline
    \hline
    \multirow{2}{*}{Bugzilla} & \texttt{Eclipse} & 27,583 & 1,447 (5.2\%) & 959 & 2.5 & 19 \\
    & \texttt{Mozilla} & 193,587 & 20,189 (10.4\%) & 10,702 & 2.9 & 151 \\
    \hline
    \multirow{2}{*}{Jira} & \texttt{Hadoop} & 14,016 & 377 (2.7\%) & 336 & 2.1 & 6\\
    & \texttt{Spark} & 9,579 & 354 (3.7\%) & 290 & 2.2 & 14 \\
    \hline
    \multirow{2}{*}{GitHub} & \texttt{Kibana} & 17,016 & 470 (2.8\%) & 388 & 2.2 & 9\\
        & \texttt{VSCode} & 62,092 & 4386 (7\%) & 2,342 &  2.9 & 51 \\
    \hline
\end{tabular}
\begin{tablenotes}
    \small
    \item Avg. BRs: average number of BRs in a bucket; Max. BRs: max number of BRs in a bucket
\end{tablenotes}
\end{threeparttable}
\end{table*}

Overall, we present the basic statistics of the six projects in Table~\ref{tab:data_six_projs}. 
We can find that different projects have different characteristics: in one bucket, \texttt{Mozilla} can have 151 BRs that are duplicates of each other, while for \texttt{Hadoop}, the maximum BRs in one bucket is only 6. 
The ranking of the total number of BRs in each ITS is Jira < GitHub < Bugzilla. 
On average, the percentage of duplicate BRs is also Jira < GitHub < Bugzilla.

%% file: sec/4_experiment.tex
\section{Empirical Settings}
\label{sec:experiments}

\subsection{Experimental Setup}
\label{sec:exp-setup}
We design experiments to answer the RQs described in Section~\ref{sec:introduction}. 
For RQ1, we focus on understanding the biases that may affect the performance of DBRD techniques. 
For RQ2, we conduct a comparison among the existing DBRD techniques on our new benchmark. 
RQ3 focuses on the DBRD techniques in practice.

\vspace{0.2cm}\noindent{\bf RQ1}: To answer RQ1, we evaluate the three best-performing DBRD techniques reported in a recent study by Rodrigues et al.~\cite{rodrigues2020soft}: \texttt{REP}~\cite{sun2011towards}, \texttt{Siamese Pair}~\cite{deshmukh2017towards}, and \texttt{SABD}~\cite{rodrigues2020soft}.
For the age bias (old vs. recent data) and state bias (initial vs. latest state), we run experiments on the BRs from Bugzilla. For the ITS bias, we run experiments on the BRs from Bugzilla, Jira and GitHub (Bugzilla vs. Jira, and Bugzilla vs. GitHub).
To analyze the impact of age bias and state bias, we pick two popular projects (\texttt{Mozilla} and \texttt{Eclipse}) which use Bugzilla as their ITS as a starting point.
We conduct controlled experiments~\cite{basili2007role, garousi2020contemporary} by varying one variable (i.e., age and state) at a time. Furthermore,  We choose \texttt{Mozilla} and \texttt{Eclipse} projects due to the long history associated with them in terms of both the number of BRs and the number of duplicates.

To investigate the impact of {\em age bias}, we evaluate the effectiveness of the three DBRD techniques on BRs from two time windows: old (2012--2014) vs. recent (2018--2020). Table~\ref{tab:rq1_age} shows the data statistics of these two time windows.

\begin{table}[t]
\footnotesize
\caption{Statistics of old (2012--2014) and recent (2018--2020) data for RQ1}
\label{tab:rq1_age} 
\centering
\begin{tabular}{|l|l|r|r|r|r|r|r|r|}
\hline
\multirow{2}{*}{\textbf{Project}} & \multirow{2}{*}{\textbf{Age}} & \multicolumn{2}{c|}{\textbf{Train}} & \multicolumn{1}{c|}{\textbf{Test}} & \multicolumn{2}{c|}{\textbf{Total}}  \\
\cline{3-7}
& & \textbf{\# BRs (\% Dup)} & \textbf{\# Dup Pairs} & \textbf{\# BRs (\% Dup)} & \textbf{\# BRs (\% Dup)} & \textbf{\# Master BRs}\\
\hline
\hline
\multirow{2}{*}{\texttt{Mozilla}} & Old  & 198,653 (9.9\%) & 35,474 & 139,502 (9.9\%) & 338,155 (9.9\%) & 21,554 \\
\cline{2-7}
& Recent & 137,886 (10.1\%) & 60,498 & 55,701 (11.2\%) & 193,587 (10.4\%) & 10,702 \\
\hline
\multirow{2}{*}{\texttt{Eclipse}} & Old  & 49,355 (5.5\%) & 4,482 & 25,021 (12.1\%) & 74,376 (7.7\%) & 3,254
\\
\cline{2-7}
& Recent & 19,607 (4.7\%) & 1,725 & 7,976 (6.5\%) & 27,583 (5.2\%) & 959  \\
\hline
\end{tabular}
\end{table}

To investigate the impact of {\em state bias}, we use \texttt{Mozilla} and \texttt{Eclipse} BRs that were submitted in 2018--2020.
For these BRs, we compare the effectiveness of the three DBRD techniques using the latest and initial states of their fields.
Among the fields we extracted for experiments, only the description cannot be changed~\cite{tu2018careful}, thus we try to recover all the other fields if changed. We recover the initial state of these fields by tracing BR's change history\footnote{An example of change history of a BR: \url{https://bugzilla.mozilla.org/show_activity.cgi?id=122876}}.
Each item in the change history of a BR describes the author, time, updated field, removed value, and added value. Table~\ref{tab:rq1_state} shows the percentage of BRs which changed their initial states.

\begin{table}[t]
\caption{The percentage of BRs changed the corresponding state in 2018--2020} \label{tab:rq1_state} 
\centering
\small
\begin{tabular}{|l|r|r|r|r|r|r|r|}
\hline
\textbf{Platform} & \textbf{Summary} & \textbf{Description} & \textbf{Product} & \textbf{Component} &  \textbf{Priority} & \textbf{Severity} & \textbf{Version} \\
\hline
\hline
\texttt{Eclipse} &  10.8\% & - & 7.8\% & 11.7\%  & 1.2\%  & 5.6\% & 8.6\% \\
\hline
\texttt{Mozilla} & 11.8\% & - & 21.4\% &  24.5\% & 24.5\% & 5.4\% & 4.2\% \\
\hline
\end{tabular}
\end{table}

To investigate the impact of {\em ITS bias}, we evaluate the three DBRD techniques on BRs (reported in 2018--2020) for three sets of projects that use Bugzilla (\texttt{Eclipse} and \texttt{Mozilla}), Jira (\texttt{Spark} and \texttt{Hadoop}) and GitHub (\texttt{Kibana} and \texttt{VSCode}) as ITS.

\vspace{0.2cm}\noindent{\bf RQ2}: To answer RQ2, other than the three techniques evaluated for answering RQ1, we add two more recent techniques, \texttt{DC-CNN}~\cite{he2020duplicate} and \texttt{HINDBR}~\cite{xiao2020hindbr}. 
These two approaches are initially designed and also evaluated as a classification task. 
In our experimental setting, all the BRs submitted within a one-year time window are duplicate BRs candidates. 
For each BR $br_i,i=1,...,m$, where $m$ is the total number of BRs in the test set. We pair it with all its candidate BRs, i.e., $(br_i, br_{i, 1}), (br_i, br_{i, 2}) ... (br_i, br_{i, k})$, where $k$ is the total number of candidate BRs of $br_i$, $br_{i,j}, j={1,...,k}$ is a candidate duplicate of $br_i$. 
The trained classification model is used to predict whether each pair is duplicate. 
The output probability value is then used to rank the possibility that each candidate BR be a duplicate one.
If two candidate BRs have the same probability value, we rank them based on their BR ids in ascending order.
We get the top-$k$ recommendations and evaluate the performance based on the recommendations. 
We evaluate the five techniques on a new benchmark dataset unaffected by the biases that are found to have a significant and substantial impact on DBRD evaluation in RQ1.

\vspace{0.2cm}\noindent{\bf RQ3}: To answer RQ3, we investigate the effectiveness of the DBRD techniques used in practice by \texttt{Mozilla} and \texttt{VSCode} projects, and compare them with the aforementioned five DBRD techniques.

\texttt{Mozilla} uses Bugzilla as its ITS, and Bugzilla implements several variants of a DBRD technique named Full-Text Search (\texttt{FTS})\footnote{\url{https://github.com/bugzilla/bugzilla/blob/5.2/Bugzilla/Bug.pm} and \url{https://github.com/bugzilla/bugzilla/blob/5.2/Bugzilla/DB.pm}}.
We study the source code of Bugzilla to identify how \texttt{FTS} works. Simply put, based on the summary input, FTS relies on a BR database and issues SQL queries to search in the database.
There are two variants of FTS: \texttt{FULLTEXT\_OR} and \texttt{FULLTEXT\_AND}. 
For the first variant, the SQL queries specify `OR' operations in full-text search; otherwise, the SQL queries specify an exact match.
We inferred the variant of the \texttt{FTS} search used by \texttt{Mozilla} (i.e., the OR variant) by trying to enter new BRs into its \texttt{FTS}.
We replicated the OR-variant by using the same SQL queries that Bugzilla uses on reading the entire summary field.

\begin{table}[t]
\caption{The number of issues per duplicate recommendation frequency by \texttt{VSCodeBot}}
\label{tab:vscodebot-data}
\centering
\begin{tabular}{|lrrrrrr|}
    \hline
    \# predictions  & 1 & 2 & 3 & 4 & 5 & Total \\
    \hline
    \# issues & 5,016 & 2,006 & 1,158 & 730 & 2,898 & 11,808 \\
    \hline
\end{tabular}
\end{table}

Another DBRD technique used in practice is \texttt{VSCodeBot}~\cite{vscodebot}, but its implementation is not publicly available, and it only works for the \texttt{VSCode} repository.
Thus, we only evaluate the five techniques on the test data from the \texttt{VSCode} project.
Table~\ref{tab:vscodebot-data} presents the number of \texttt{VSCode} issues that have potential duplicate recommendations by \texttt{VSCodeBot} made in 2018-2020.
Among the 62,092 \texttt{VSCode} BRs in our dataset, 11,808 BRs got potential duplicate recommendations.
\texttt{VSCodeBot} only recommends up to five recommendations for an issue.

\subsection{Evaluation Metrics}
\label{sec:metrics}
We evaluate the effectiveness of DBRD tools by calculating Recall Rate@$k$ (RR@$k$).
The evaluation strategy is consistent with the previous works for the DBRD task~\cite{deshmukh2017towards,runeson2007detection,wang2008approach,sun2010discriminative,amoui2013search}, i.e., only RR@$k$ is used.
Note that another metric that has been used in a few DBRD research papers~\cite{sun2011towards,rodrigues2020soft} is Mean Average Precision (MAP). 
MAP concerns the position of the ground truth master BR in each prediction. A higher MAP means that for each BR in the test set, the model can return the ground truth master at a higher place in the ranked result list. The primary reason that prior works, as well as our work, do not consider MAP is that it does not simulate the real usage scenario: In real use, it is unlikely developers would frequently check the recommendations after 10 BRs. For example, prior work on understanding practitioners’ expectations on automated fault localization~\cite{kochhar2016practitioners} has shown that nearly all the respondents (close to 98\%) are unwilling to inspect more than ten program elements to find the faulty code. Another supporting evidence is for \texttt{VSCodeBot}~\cite{vscodebot} used in the Microsoft Visual Studio Code repository, the largest number of duplicate issue recommendations is 5.

\begin{figure}[t]
    \centering
    \includegraphics[width=0.7\textwidth]{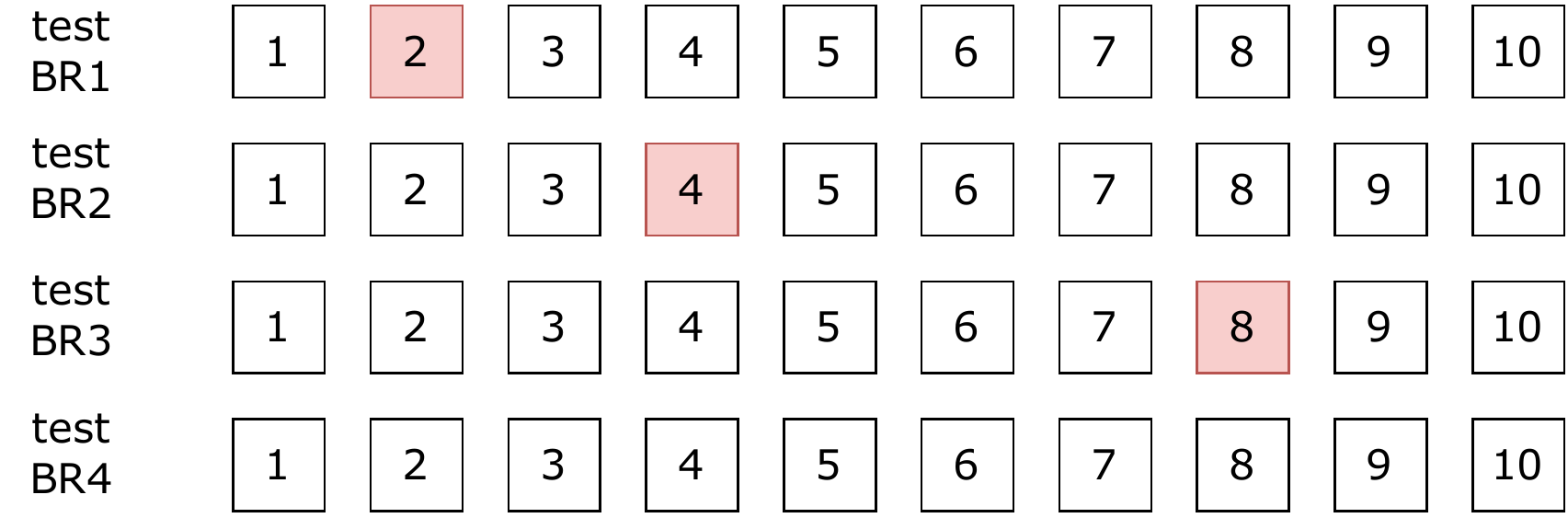}
    \caption{Examples of the predictions in the top-10 positions for 4 test BRs.}
    \label{fig:recall-rate}
\end{figure}

\textit{RR@$k$} is defined as follows:
\begin{equation*}
    RR@k = \frac{n_k}{m},
\end{equation*}
where $n_k$ is the number of duplicate BRs in the test set whose bucket has been found in the top-$k$ positions;
$m$ is the total number of duplicate BRs in the test set.

For example, in Figure~\ref{fig:recall-rate}, consider a project with four BRs in the test set. We show the top-10 predictions for the four test BRs. If the corresponding prediction is correct, the box is highlighted in red. In this example, for the test BR1, BR2, BR3, the successful prediction is in the 2nd, 4th, and 8th predictions, respectively. For the test BR4, all the top-10 predictions are wrong. The RR@$k$ in this dataset would be: RR@$k(k=1)$ = $0$,
RR@$k(k=2,3)$=$1/4$=$0.25$,
RR@$k(k=4,5,6,7)$=$2/4$=$0.5$, and 
RR@$k(k=8,9,10)$=$3/4$=$0.75$.

The process of finding a duplicate BR's master BR is essentially equivalent to finding the bucket to which it belongs.
In the prediction stage, for each BR in the test dataset, a DBRD technique queries its candidate BRs and returns a possible duplicate BR list.
In the evaluation stage, we group the BRs in the list returned by a DBRD technique into \textit{buckets} list to calculate the recall rate of the predictions.
This evaluation strategy is consistent with Rodrigues et al.~\cite{rodrigues2020soft}.

\begin{figure}[t]
    \centering
    \includegraphics[width=0.8\textwidth]{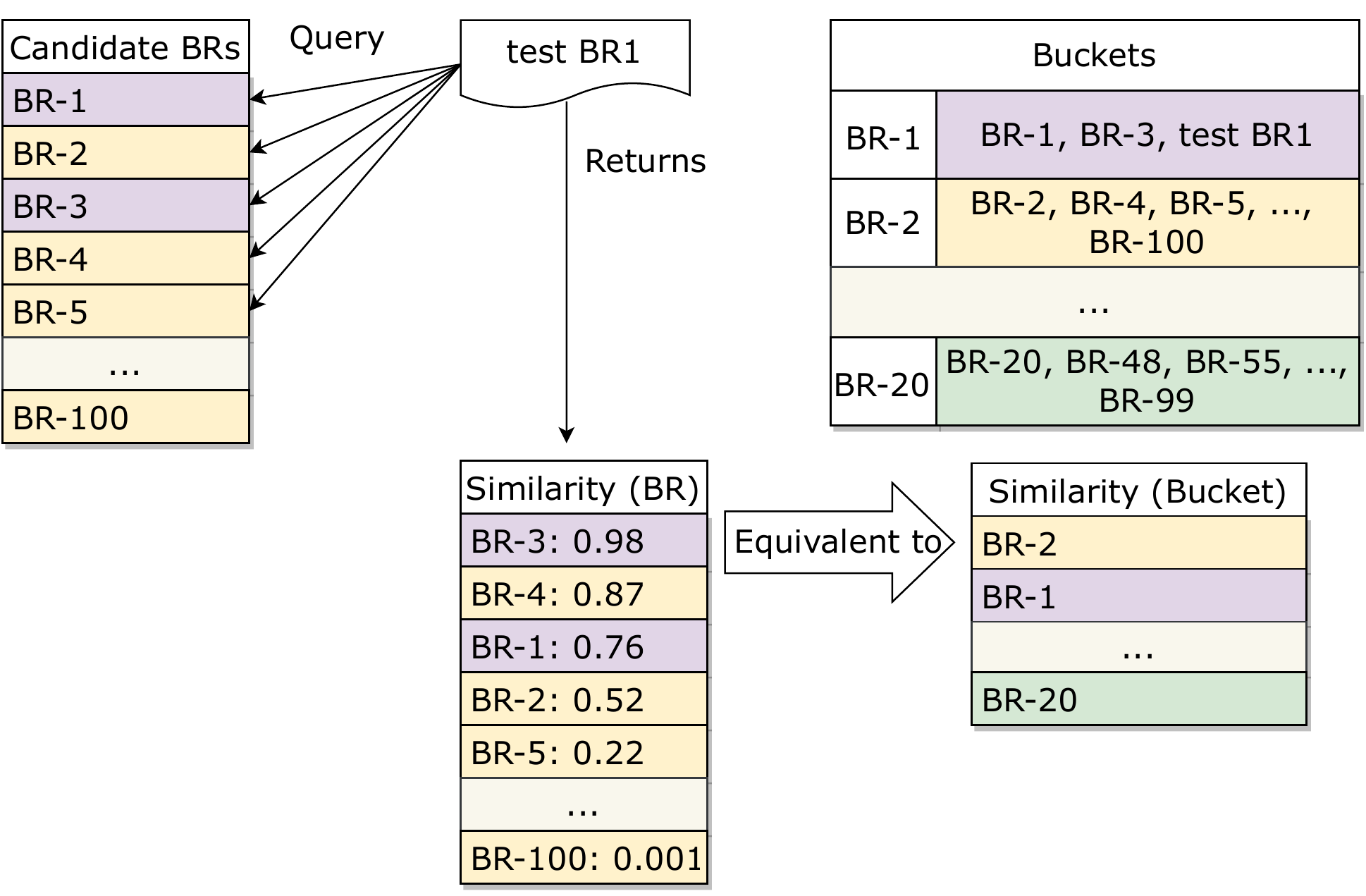}
    \caption{The workflow of retrieving the correct bucket.}
    \label{fig:workflow}
\end{figure}

In this paragraph, we elaborate on how $n_k$ in RR@$k$ is calculated on the example in Figure~\ref{fig:recall-rate}. 
There are four BRs in the test set. 
To find the BRs that are duplicates of a test BR (i.e., the bucket they belong to), instead of only calculating the similarity score of the test BR with master BRs, we compare the test BR with all the BRs in buckets. Like the prior works~\cite{sun2011towards,rodrigues2020soft}, we also use the highest score among the test BR with all the BRs in the candidate bucket as the similarity score between the BR with the candidate bucket. 
We illustrate a concrete example as shown in Figure~\ref{fig:workflow}. We have 100 BRs: $\{BR\mhyphen 1, BR\mhyphen 2, BR\mhyphen 3, ..., BR\mhyphen 100\}$ and they belong to 20 buckets BR-1 : $\{BR\mhyphen1, BR\mhyphen3, test BR1\}$, BR-2 : $\{BR\mhyphen2, BR\mhyphen4, BR\mhyphen5, ..., BR\mhyphen100\}$, BR-3: $\{BR\mhyphen3,...\}$, ... BR-20: $\{BR\mhyphen20, BR\mhyphen48, BR\mhyphen55, ..., BR\mhyphen99\}$. 
We represent a bucket as a dictionary: the key is the master BR, and the value is its duplicate BRs and itself. 
Thus, we use the master BR id to refer to the bucket. Given $test BR1$, each model will return a list of BRs sorted by the likelihood of being a duplicate of test BR1. The returned rank list is: [(BR-3, 0.98), (BR-4, 0.87), (BR-1, 0.76), (BR-2, 0.52), (BR-5, 0.22), ... (BR-100, 0.001)]. 
The first part of each tuple is the BR id, and the second part is the similarity score. According to the ground truth information, test BR1 belongs to the bucket BR-1: $\{BR\mhyphen1, BR\mhyphen3, test BR1\}$. 
As test BR1 is in the same bucket as BR-1 and BR-3, the highest similarity scores, i.e., (BR-3, 0.98), will be the similarity score between test BR1 with the bucket BR-1 : $\{BR\mhyphen1, BR\mhyphen3, test BR1\}$. So in this example, this model makes a successful prediction at the second position and we can get $n_k=1(k = 2)$. 
With the same strategy, we can calculate the $n_k$ in the rest three BRs and get RR@$k(k = 2)$=$1/4$=$0.25$. 

%% file: sec/5_result.tex
\section{Empirical Results and Analyses}
\label{sec:results}
\subsection{RQ1. How significant are the potential biases on DBRD techniques?}

Due to the page constraint, we only present the statistical test results in Table~\ref{tab:rq1_significance}.
The detailed results are available in our online appendix.\footnote{\url{https://github.com/soarsmu/TOSEM-DBRD}}

\vspace{0.2cm}\noindent{\bf Age Bias:} We run the Mann-Whitney U test~\cite{gehan1965generalized} on the following null hypothesis for each pair of DBRD approach and project:

$H_{0.1}$: There is no significant difference in RR@$k$ on old issues and recent issues.

We also compute Cliff's delta (d) effect size \cite{romano2006appropriate}.
As we have six $p$-values on the same hypothesis, we also run Bonferroni correction~\cite{weisstein2004bonferroni}, and the significance level $\alpha$ becomes 0.0083.
We find that the $p$-value is < 0.0083 for all approach-project pairs except one case (and thus we can reject the null hypothesis) and the effect size is large.
We contend that the age of data significantly affects DBRD performance.

\vspace{0.2cm}\noindent{\bf State Bias:} Similar to the prior bias, we run the Mann-Whitney U test on the following null hypothesis for each pair of the DBRD approach and project:

$H_{0.2}$: There is no significant difference in RR@$k$ on the initial state and recent state within an issue.

We also compute Cliff's delta effect size. With Bonferroni correction, the significance level $\alpha$ is 0.0083. 
We find that the $p$-value is > 0.0083 for all approaches on projects with the initial and latest state (and thus we cannot reject the null hypothesis).
We contend that the state does not make a significant difference in DBRD performance.

\vspace{0.2cm}\noindent{\bf ITS Bias:} we also run the Mann-Whitney U test on the following null hypothesis for each approach on Bugzilla vs. Jira, and Bugzilla vs. GitHub data:

$H_{0.3}$: There is no significant difference in RR@$k$ on Bugzilla issues and other ITS issues.

We also compute Cliff's delta effect size.
We find that the $p$-value is < 0.0083 (with Bonferroni correction) for all approaches and the effect size is large (except for one case), and thus we can reject the null hypothesis.
Thus, we contend that ITS plays an important role in DBRD technique performance.

\begin{table}[t]
\caption{Mann-Whitney-U with Cliff's Delta Effect Size $|d|$ on RQ1} \label{tab:rq1_significance} 
\centering
\small
\begin{tabular}{|l|l|lrr|}
    \hline
    \textbf{Bias}
    & \textbf{Approach}
    & \textbf{Data}
    & \textbf{$p$-value}
    & \textbf{$|d|$} \\
    \hline
    \hline
    \multirow{6}{*}{Age} & \multirow{2}{*}{\texttt{REP}} 
    & \texttt{Eclipse}  &0.003 & 0.78 (large) \\
    &  & \texttt{Mozilla} & 0.005 & 0.72 (large)  \\
    \cline{2-5}
& \multirow{2}{*}{\texttt{Siamese Pair}}  & \texttt{Eclipse} & 
$<0.001$ & 1 (large)  \\
    &  & \texttt{Mozilla} & 0.003 & 0.76 (large)  \\
        \cline{2-5}
& \multirow{2}{*}{\texttt{SABD}} 
    & \texttt{Eclipse} & 0.001 & 0.82 (large)  \\
    &  & \texttt{Mozilla} & 0.012 & 0.66 (large)  \\
    \hline
    \hline
    \multirow{6}{*}{State} &
    \multirow{2}{*}{\texttt{REP}} 
    & \texttt{Eclipse}  & 0.105  & 0.44 (medium) \\
    &  & \texttt{Mozilla} &  0.190 & 0.36 (medium) \\
    \cline{2-5}
& \multirow{2}{*}{\texttt{Siamese Pair}} & \texttt{Eclipse}
& 0.063 & 0.5 (large) \\
    &  & \texttt{Mozilla} & 0.190 & 0.36 (medium) \\
        \cline{2-5}
    & \multirow{2}{*}{\texttt{SABD}}
    & \texttt{Eclipse} & 0.315  & 0.28 (small) \\
    &  & \texttt{Mozilla} & 0.315  & 0.28 (small) \\
    \hline
    \hline
    \multirow{6}{*}{ITS} &
    \multirow{2}{*}{\texttt{REP}}
    & Jira & 0.056 & 0.36 (medium) \\
    &  & GitHub &  $<0.001$  & 0.66 (large) \\
        \cline{2-5}
& \multirow{2}{*}{\texttt{Siamese Pair}}
    & Jira  & $<0.001$  & 1 (large) \\
    &  & GitHub & $<0.001$ & 0.97 (large) \\
    \cline{2-5}
    & \multirow{2}{*}{\texttt{SABD}}
    & Jira  & $<0.001$ & 0.97 (large)  \\
    & & GitHub  & $<0.001$ & 0.77 (large)\\
\hline
\end{tabular}
\end{table}

\begin{tcolorbox}[
    left=0pt, right=0pt,
    top=0pt, bottom=0pt,
    boxrule=0pt, frame empty]
\textbf{Answer 1:} \textit{{\em Age bias} has a statistically significant impact (with large effect size) on the evaluation of DBRD techniques in all but one cases. {\em State bias} does not have a statistically significant impact. {\em ITS bias} has a statistically significant impact (with large effect size) on all but one case.}
\end{tcolorbox}

\subsection{RQ2. How do the state-of-the-art DBRD research tools perform on recent data?}

\begin{figure*}
    \begin{minipage}[b]{0.5\textwidth}
    \centering
    \includegraphics[width=\textwidth]{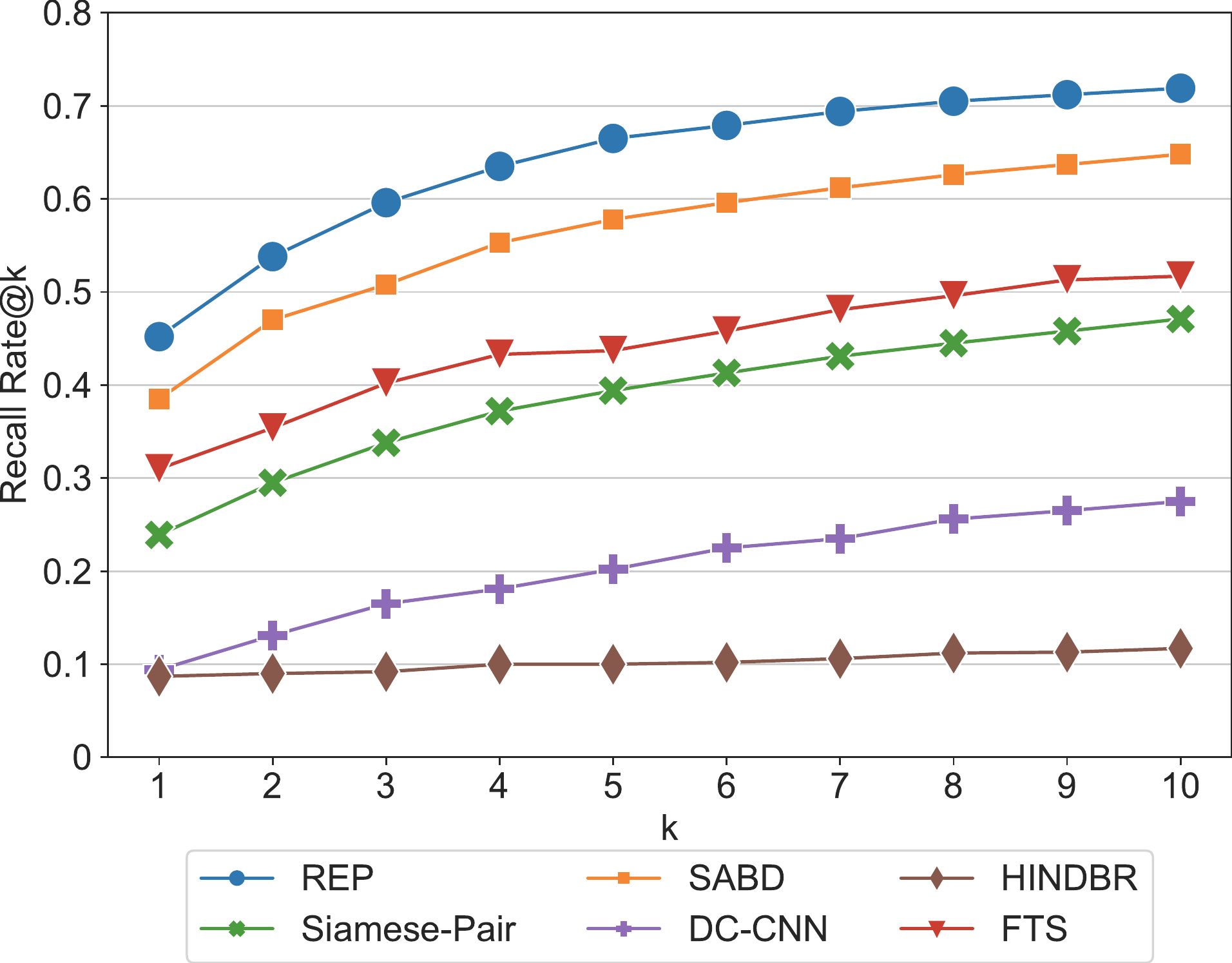}\\
    \subcaption{\texttt{Eclipse}}
    \end{minipage}%
    \begin{minipage}[b]{0.5\textwidth}
    \centering
    \includegraphics[width=\textwidth]{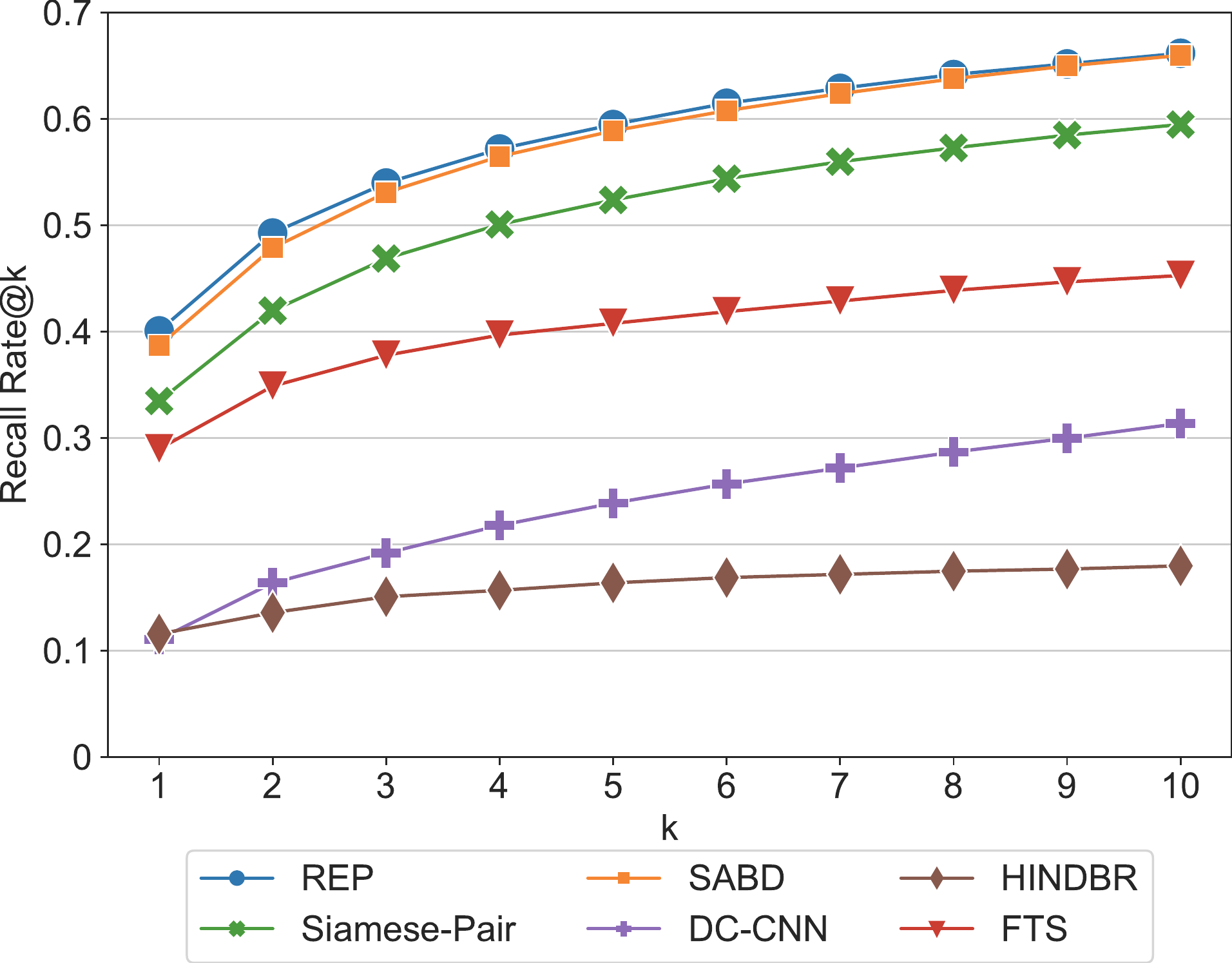}\\
    \subcaption{\texttt{Mozilla}}
    \end{minipage}
\\
    \begin{minipage}[b]{0.5\textwidth}
    \centering
    \includegraphics[width=\textwidth]{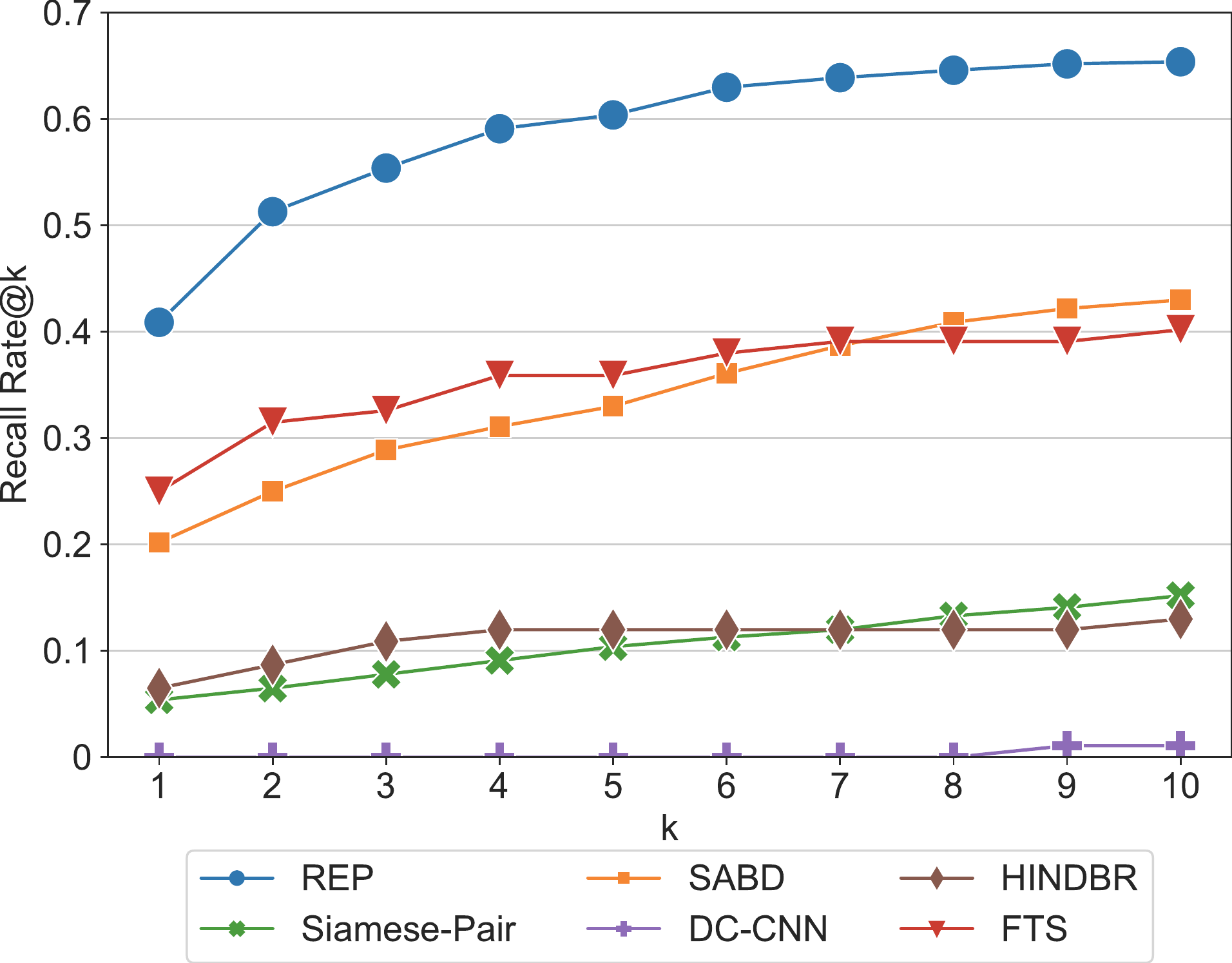}\\
    \subcaption{\texttt{Hadoop}}
    \end{minipage}%
    \begin{minipage}[b]{0.5\textwidth}
    \centering
    \includegraphics[width=\textwidth]{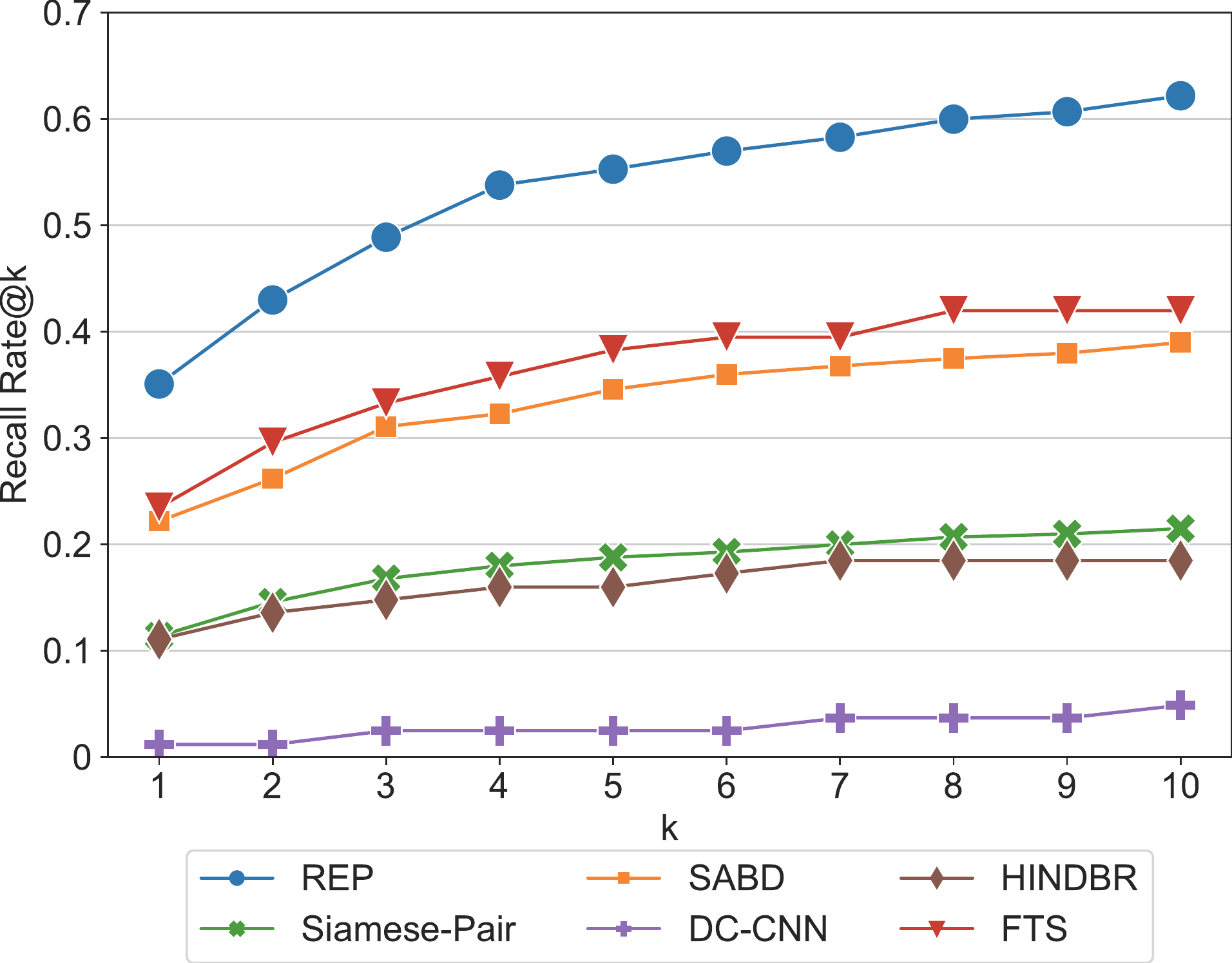}\\
    \subcaption{\texttt{Spark}}
    \end{minipage}
\\
    \begin{minipage}[b]{0.5\textwidth}
    \centering
    \includegraphics[width=\textwidth]{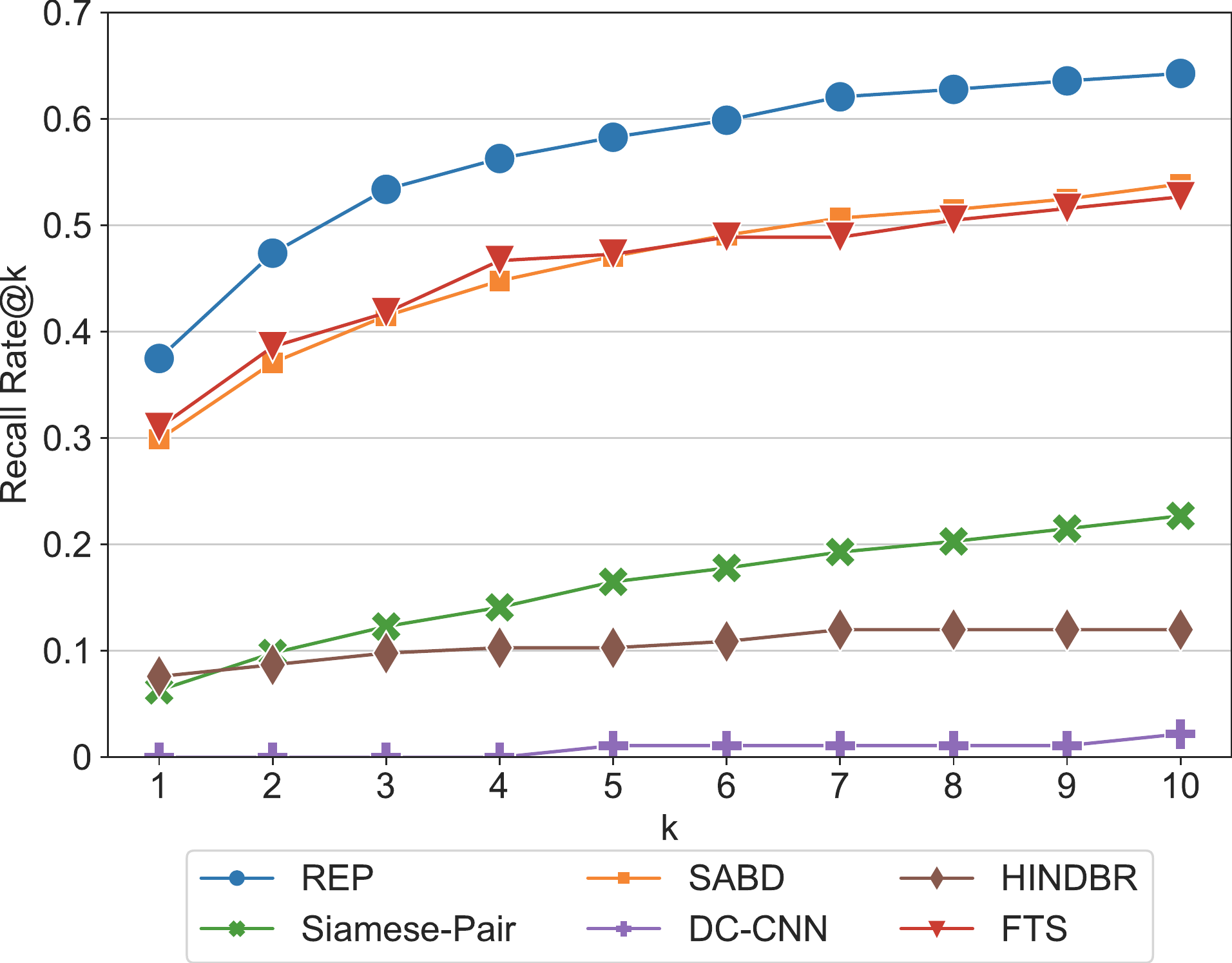}\\
    \subcaption{\texttt{Kibana}}
    \end{minipage}%
    \begin{minipage}[b]{0.5\textwidth}
    \centering
    \includegraphics[width=\textwidth]{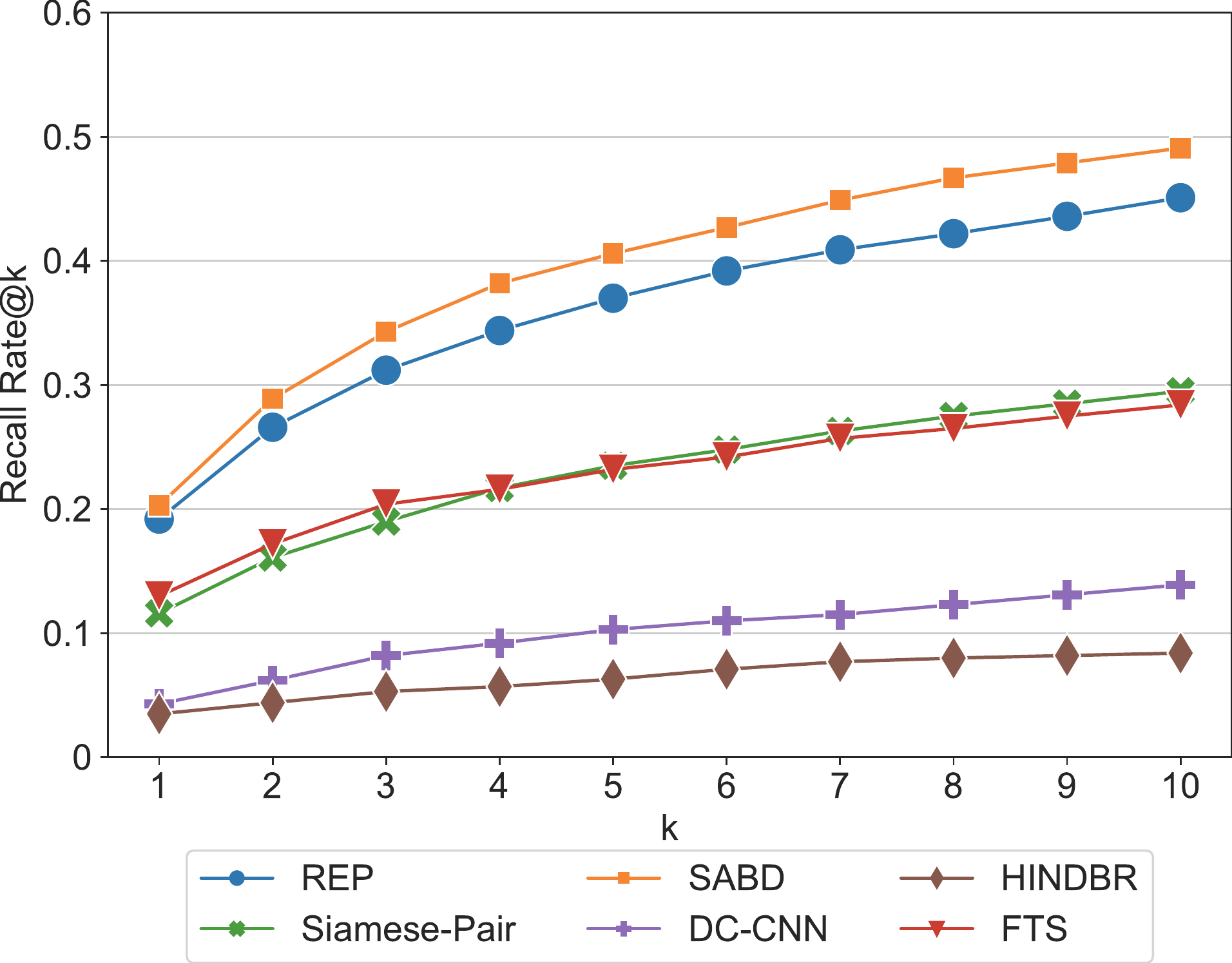}\\
    \subcaption{\texttt{VSCode}}
    \end{minipage}
\caption{Recall Rate@k in the test data of \texttt{Eclipse}, \texttt{Mozilla}, \texttt{Hadoop}, \texttt{Spark}, \texttt{Kibana}, and \texttt{VSCode}}
\label{fig:results}
\end{figure*}

\begin{table*}[t]
\caption {Statistics of training and testing data}
\label{tab:our_dataset} 
\centering
\small
\begin{tabular}{|l|l|rr|r|r|}
    \hline
    \multirow{2}{*}{\textbf{ITS}} & \multirow{2}{*}{\textbf{Project}} & \multicolumn{2}{c|}{\textbf{Train}} & \multicolumn{1}{c|}{\textbf{Test}}  & \multicolumn{1}{c|}{\textbf{Total}}  \\
    \cline{3-6}
& & \# BRs (\% Dup) & \# Dup Pairs & \# BRs (\% Dup) & \# BRs (\% Dup) \\
    \hline
    \hline
    \multirow{2}{*}{Bugzilla} & \texttt{Eclipse} & 19,607 (4.7\%) & 1,725 & 7,976 (6.5\%) &  27,583 (5.2\%)\\
    & \texttt{Mozilla} & 137,886 (10.1\%)  & 35,474  & 55,701 (11.2\%) & 193,587 (10.4\%) \\
    \hline
    \multirow{2}{*}{Jira} & \texttt{Hadoop} & 10,276 (2.8\%) &  328 &  3,740 (2.5\%) & 14,016 (2.7\%) \\
    & \texttt{Spark} & 6,738 (4\%) & 414 & 2,841 (3\%) & 9,579 (3.7\%) \\
    \hline
    \multirow{2}{*}{GitHub} & \texttt{Kibana} & 9,849 (2.9\%)  & 376 & 7,167 (2.6\%) & 17,016 (2.8\%) \\
        & \texttt{VSCode} & 40,801 (7.2\%) & 9,008 & 21,291 (6.8\%) & 62,092 (7\%)\\
    \hline
\end{tabular}
\end{table*}

Based on the findings from RQ1, we evaluate the existing DBRD techniques on a new benchmark that omits {\em age bias} and {\em ITS bias}.
Our benchmark contains the recent three-year BRs extracted from Bugzilla, Jira, and GitHub.
We build this dataset as described in Section~\ref{sec:data}.
Table~\ref{tab:our_dataset} shows the statistics of the training and testing data.
Figure~\ref{fig:results} shows RR@$k$ from the 5 approaches on our dataset. 
The x-axis denotes $k$ values from 1 to 10 while the y-axis shows RR@$k$.
Each approach is highlighted with a different shape and color.
For all datasets, \texttt{REP} outperforms the other approaches except for \texttt{Mozilla} and \texttt{VSCode}. 
On average, \texttt{REP} outperforms \texttt{SABD} by $22.3\%$ in terms of RR@$10$ across 6 project data.
As presented in Table~\ref{tab:our_dataset}, both \texttt{Mozilla} and \texttt{VSCode} have the largest number of BRs and duplicates BRs.
For these two projects, \texttt{SABD} shows comparable results with \texttt{REP} and even outperforms on \texttt{VSCode} dataset by $9\%$ in terms of RR@$10$.
\texttt{Siamese Pair}, \texttt{HINDBR}, and \texttt{DC-CNN} show fluctuating results.
\texttt{Siamese Pair} presents higher RR@$k$ values on \texttt{Eclipse}, \texttt{Mozilla}, and \texttt{VSCode}, while it demonstrates lower RR@$k$ than \texttt{HINDBR} and \texttt{Hadoop} on \texttt{Spark} and \texttt{Kibana}. 

\begin{table}[!ht]
\caption {Investigation of which component benefits \texttt{REP}}
\label{tab:component} 
    \centering
    \small
    \begin{tabular}{|l|l|l|l|l|l|l|l|}
    \hline
        \multirow{2}{*}{\textbf{RR@$k$}} & \multirow{2}{*}{\textbf{All}} &   \multicolumn{5}{c|}{\textbf{w/o}}\\
        \cline{3-7}
        & & \textbf{\texttt{description}} & \textbf{\texttt{short\_desc}} & \textbf{\texttt{product}} & \textbf{\texttt{component}} & \textbf{\texttt{priority}} \\ 
        \hline
        \hline
        1 & 0.460 & 0.327 & 0.350 & 0.456 & 0.458 & 0.450 \\ \hline
        2 & 0.544 & 0.415 & 0.458 & 0.527 & 0.554 & 0.540 \\ \hline
        3 & 0.610 & 0.456 & 0.494 & 0.575 & 0.598 & 0.602 \\ \hline
        4 & 0.646 & 0.490 & 0.510 & 0.617 & 0.633 & 0.637 \\ \hline
        5 & 0.673 & 0.515 & 0.533 & 0.644 & 0.658 & 0.665 \\ \hline
        6 & 0.690 & 0.531 & 0.552 & 0.662 & 0.663 & 0.679 \\ \hline
        7 & 0.704 & 0.544 & 0.569 & 0.671 & 0.671 & 0.694 \\ \hline
        8 & 0.706 & 0.556 & 0.579 & 0.681 & 0.683 & 0.706 \\ \hline
        9 & 0.715 & 0.565 & 0.600 & 0.687 & 0.683 & 0.712 \\ \hline
        10 & 0.721 & 0.573 & 0.613 & 0.698 & 0.692 & 0.717 \\ \hline
    \end{tabular}
\end{table}

\textbf{Component Analysis.} As shown in Figure~\ref{fig:results}, \texttt{REP} is the winner in 5 out of the 6 datasets and it takes advantage of the information from most of the fields.
Thus, we investigate \texttt{REP} to understand the contribution of each component in DBRD. \texttt{REP} initializes the weights of the textual features higher than the rest of the features. 
Additionally, the initial weight of the summary is also higher than the weight of the description. 
Besides, among the total 19 parameters tuned by gradient descent, 14 parameters relate to the textual fields. 
Based on the observation, our assumption is that textual fields are the most crucial components among all the fields considered. 
To further verify our hypothesis, we investigate the contributions of each field value in \texttt{REP}. We run \texttt{REP} on the Eclipse dataset and each time, set a certain field value as empty. 
Table 10 demonstrates the performances when we set each field as empty. We can find that \texttt{REP} performs the best when all the information is present. When the values of the fields of severity, priority, product, and component are left empty, the performance is similar to the result when all the information is considered. The RR@$k$ is decreased at most 4\%. However, when textual information is absent, we can observe that the RR@$k$ decreased at most by 21\%. 
It indicates that textual information plays a more important role than categorical information in DBRD.

\textbf{Implications.}
Based on Figure~\ref{fig:results}, when k = 5, the best performing approach can reach RR@$k$ = 0.4-0.6.
It indicates that a model can successfully recommend the duplicate BR in the first five positions in 40\%-60\% of the cases. In other words, in the rest 40\%-60\% of the cases, the model fails to recommend a duplicate in the first five positions.
Among all the projects in our dataset, 2.7\% - 10\% BRs are duplicates. 
A technique that has the RR@$k$ of 0.4 - 0.6 means it can help eliminate at most 6\% duplicate BRs of all the BRs.
Thus, it saves considerable costs and human labor. Besides, considering the large number of candidate BRs, successfully recommending the correct master BR in the first five positions is a challenging task by itself.

\begin{figure*}[t]
    \begin{minipage}[b]{0.4\textwidth}
    \centering
    \includegraphics[width=\textwidth]{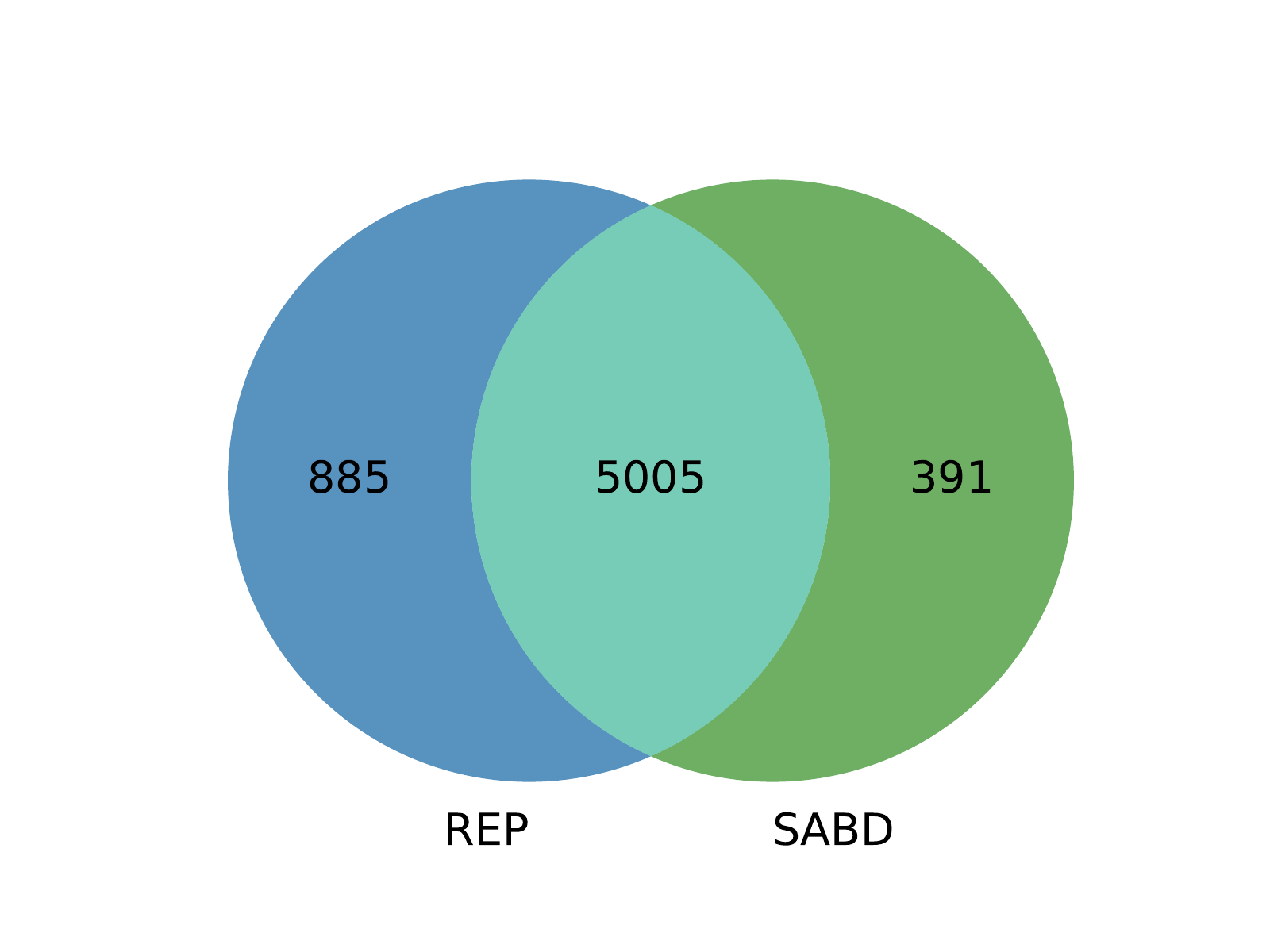}\\
    \subcaption{\texttt{REP} vs. \texttt{SABD}}
    \end{minipage}%
    \begin{minipage}[b]{0.4\textwidth}
    \centering
    \includegraphics[width=\textwidth]{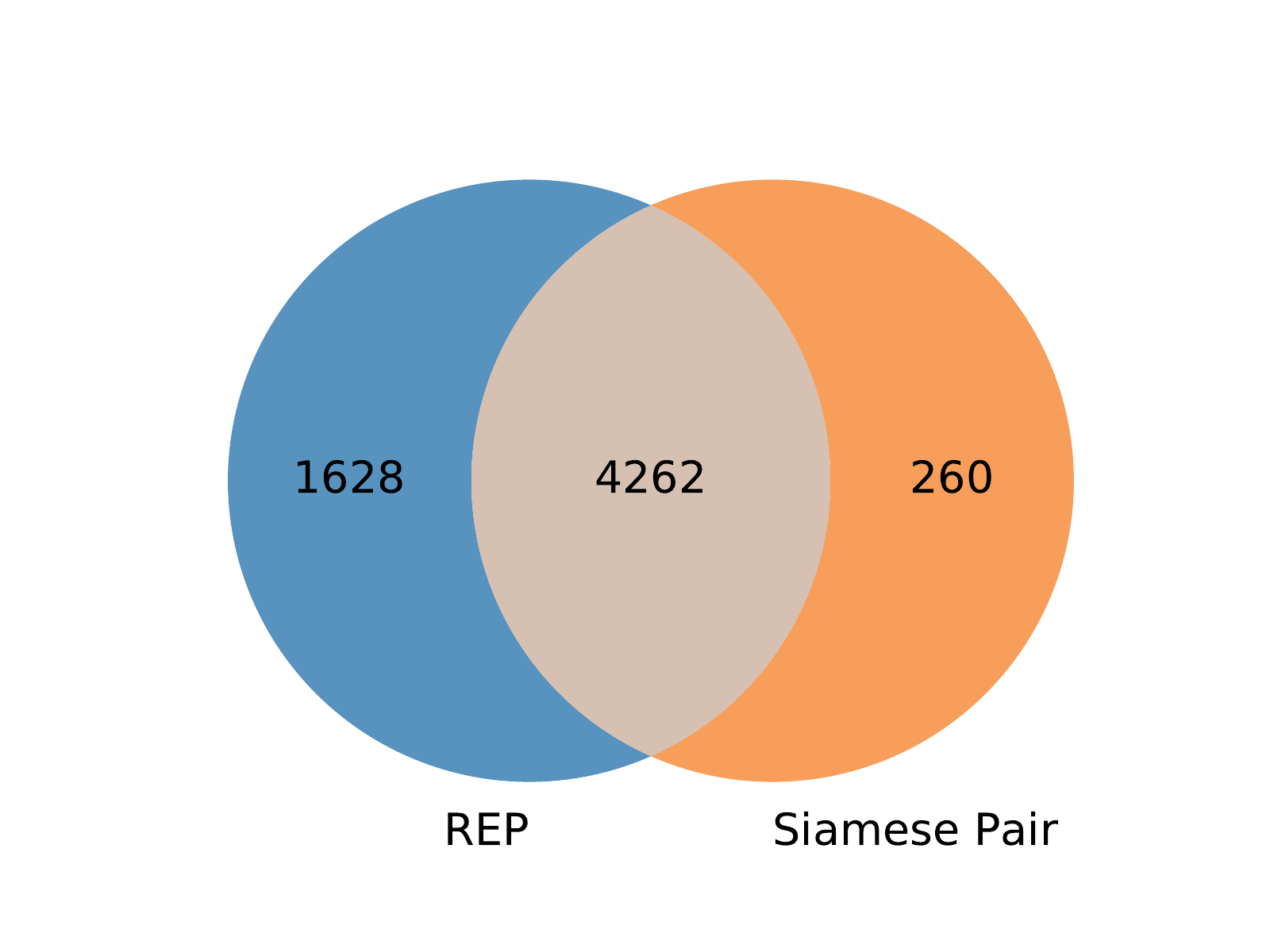}\\
    \subcaption{\texttt{REP} vs. Siamese Pair}
    \end{minipage}%
\\
    \begin{minipage}[b]{0.4\textwidth}
    \centering
    \includegraphics[width=\textwidth]{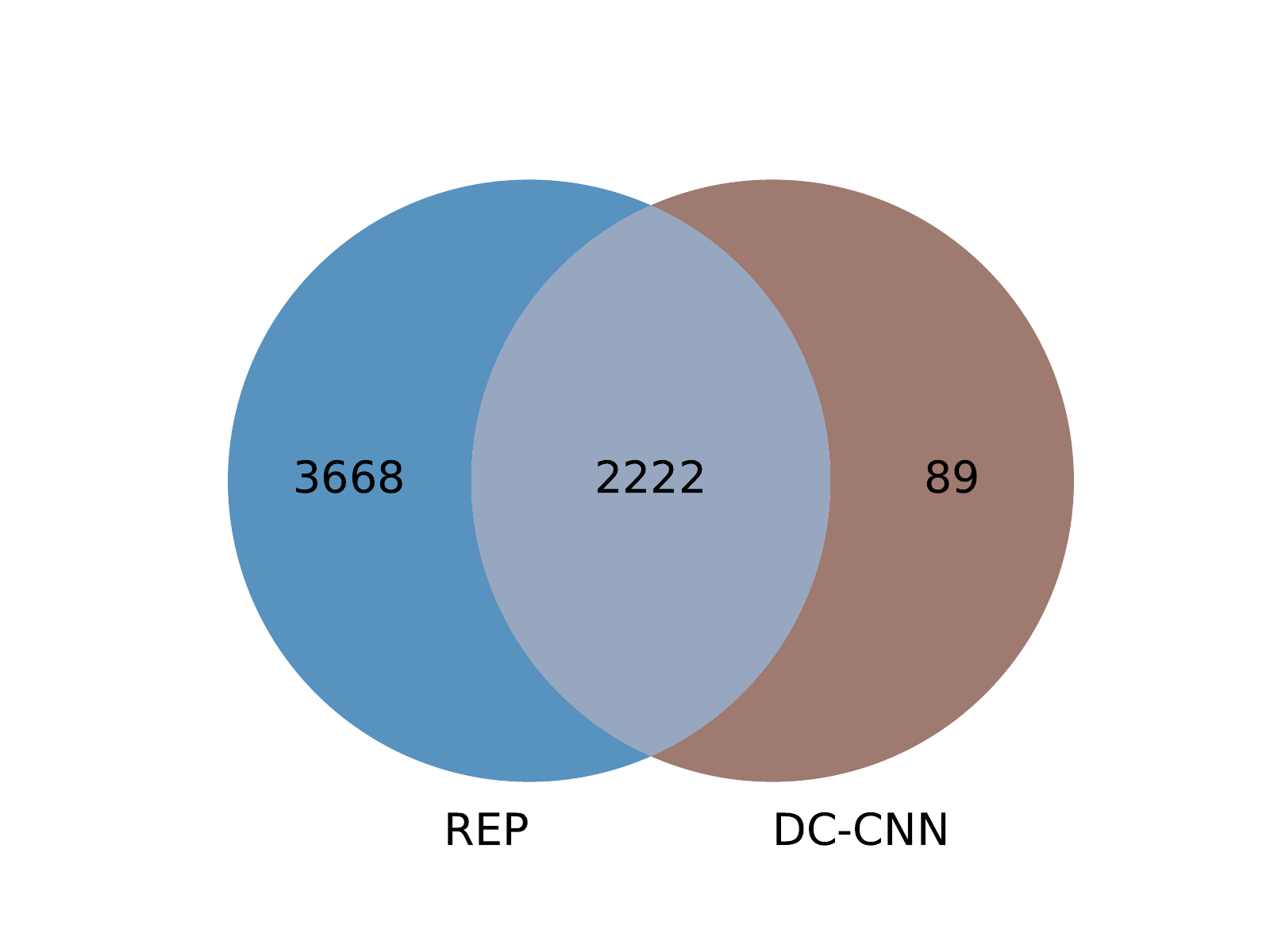}\\
    \subcaption{\texttt{REP} vs. DC-CNN}
    \end{minipage}%
    \begin{minipage}[b]{0.4\textwidth}
    \centering
    \includegraphics[width=\textwidth]{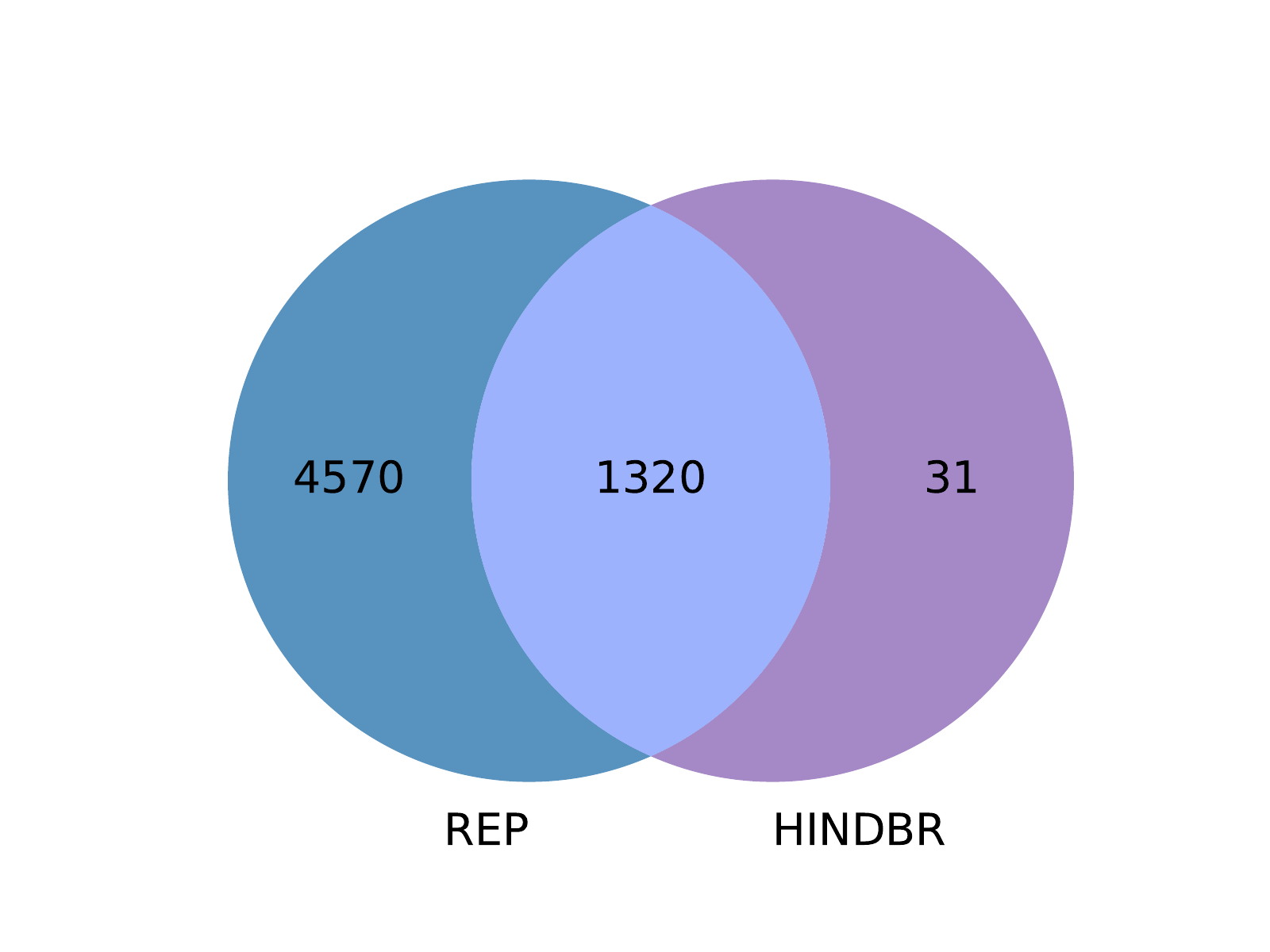}\\
    \subcaption{\texttt{REP} vs. HINDBR}
    \end{minipage}%
\caption{\texttt{REP} compared to the other four approaches in terms of successful predictions}
\label{fig:venn}
\end{figure*}

Overall, the experiment results show that it is promising to directly adopt certain DBRD approaches designed based on Bugzilla data to other ITS data, such as \texttt{REP}. However, all the approaches demonstrate relatively poor performance in the \texttt{VSCode} dataset. 
Besides, we find that deep learning-based approaches are less robust than simpler approaches. (1) \texttt{REP}, which only includes handcrafted features and parameters, is considered as a simpler approach compared to deep learning approaches. \texttt{REP} demonstrated similar performance in both \texttt{Kibana} and \texttt{Mozilla}. However, these two datasets differ in three aspects, i.e., the ITSs, the number of BRs, and the duplicate BR rate: \texttt{Mozilla} contains categorical information while \texttt{Kibana} does not, and \texttt{Mozilla} has 11 times more BRs and nearly 4 times higher duplicate BR rate than \texttt{Kibana}. (2) Since \texttt{SABD} and \texttt{Siamese Pair} show better performances in Bugzilla data than in Jira and GitHub, it implies that it may not be ideal for applying the current deep learning-based DBRD techniques that are specifically designed for one ITS to detect duplicate BRs in different ITSs. (3) \texttt{Eclipse}, \texttt{Mozilla}, and \texttt{VSCode} are the largest projects in terms of the number of BRs and the duplicate BR rate. Deep learning-based \texttt{SABD} can achieve similar or better performance than \texttt{REP}. It infers that for deep learning-based approaches, they favor more training data than simpler approaches. (4) No one can win every battle. Although \texttt{REP} is the best performer in five of the six datasets, it loses to \texttt{SABD} in the \texttt{VSCode} dataset. It suggests that no one design is better than the rest all the time, the performance of DBRD techniques could be subject to the dataset characteristics. Furthermore, we draw the Venn diagrams (Figure~\ref{fig:venn}) to demonstrate that each approach can predict duplicates that the best performer cannot.

\begin{tcolorbox}[
    left=0pt, right=0pt,
    top=0pt,
    bottom=0pt,
    boxrule=0pt, frame empty]
\textbf{Answer 2:} \textit{\texttt{REP} achieves the comparative results as \texttt{SABD} on the two largest projects and it works well on smaller projects.
\texttt{REP} demonstrates promising results, especially on small projects.
On the other hand, \texttt{SABD} shows better performance on large projects.}
\end{tcolorbox}

\subsection{RQ3. How do the DBRD techniques in research compare to those in practice?}
\vspace{0.2cm}
\textbf{Full-Text Search}
Line \texttt{FTS} in Figure~\ref{fig:results} shows the RR@$k$ (for $k$=[1...10]) across the six projects in comparison with the results for the research tools.
We find that \texttt{FTS} is a competitive baseline.
It can outperform \texttt{HINDBR} and \texttt{DC-CNN} for all the projects, and it can also outperform \texttt{Siamese Pair} on four out of six projects.
In comparison with \texttt{SABD} (the second-best performing baseline), \texttt{FTS} can achieve similar performance for three out of the six projects. 
Still, \texttt{FTS} performs worse than \texttt{REP} on all projects.
The best performing research tool (\texttt{REP}) can outperform FTS by $22.1\%$ to $62.7\%$ in terms of RR@$10$.

\begin{figure}
    \centering
    \includegraphics[width=0.5\textwidth]{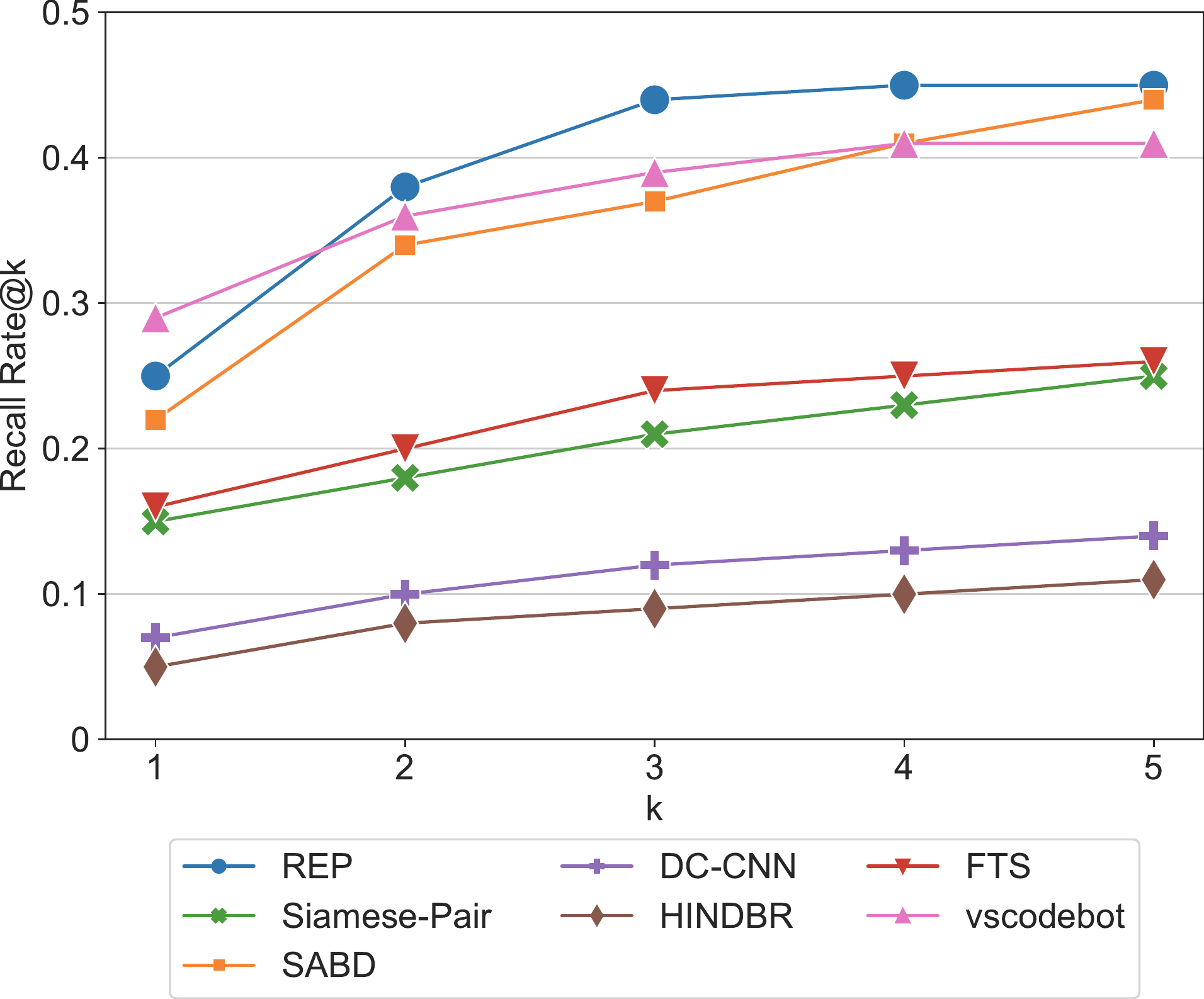}
    \caption{Recall Rate@$k$ comparing the tools in research and in practice on the VSCode data}
    \label{fig:rq3_vscodebot}
\end{figure}

\vspace{0.2cm}\noindent{\bf \texttt{VSCodeBot}} We take the intersection of \texttt{VSCode} BRs shown in Table~\ref{tab:vscodebot-data} that are truly duplicate BRs and appear in our test set (described in Table~\ref{tab:our_dataset}). We then use this data to evaluate the performance of the various DBRD approaches. Figure~\ref{fig:rq3_vscodebot} shows the RR@$k$ for each DBRD approach.
As \texttt{VSCodeBot} recommends up to five potential duplicate issues, we present RR@$k$, and the $k$ ranges from 1 to 5.
As shown in the figure, the two research tools, i.e., \texttt{REP}, \texttt{SABD}, achieved better performance than \texttt{VSCodeBot}, in terms of RR@$5$.
Meanwhile, \texttt{Siamese Pair} shows a similar result compared to \texttt{VSCodeBot} in terms of RR@$5$. 
\texttt{FTS} which is also adopted in practice also shows worse results than \texttt{VSCodeBot}.

\textbf{Implications.} Since \texttt{FTS} is based on exact word matching, the relatively good performance of \texttt{FTS} indicates that many duplicate BRs are more likely to carry the same words in BR titles. It also indicates the important role of textual information. \texttt{FTS} shows poor performance in the \texttt{VSCode} dataset, it shows that the duplicate relationship in \texttt{VSCode} dataset cannot be simply decided based on the words used in BR titles.
However, \texttt{SABD} and \texttt{REP} achieve comparable performance as \texttt{VSCodeBot} on the data we investigated, which indicates it is promising to deploy research tools in practice.

\vspace{0.3cm}
\begin{tcolorbox}[left=0pt, right=0pt, top=0pt, bottom=0pt, boxrule=0pt, frame empty]
\textbf{Answer 3:} \textit{\texttt{FTS} outperforms \texttt{HINDBR} and \texttt{DC-CNN} on all project data, and achieves competitive performance with \texttt{SABD} on three project data. On \texttt{VSCode} BRs, \texttt{REP} and \texttt{SABD} performed better than \texttt{VSCodeBot}. The best performing research tool (\texttt{REP}) increases the performance of \texttt{FTS} by $46.1\%$ in terms of average RR@$10$, and the performance of \texttt{VSCodeBot} by $9.8\%$ in terms of RR@$5$, respectively.}
\end{tcolorbox}

%% file: sec/6_discussion.tex
\section{Discussion}
\label{sec:discussion}

Based on our empirical results, this section shares our insights to benefit future research on DBRD.

\subsection{Age/State Bias in an Alternative ITS}

\begin{table}[t]
    \centering
    \small
    \caption{Age and state bias in an alternative ITS.}
    \begin{tabular}{|c|c|l|r|r|}    
    \hline
    \textbf{Bias} & \textbf{Data} & \textbf{Approach} & \textbf{$p$-value} & \textbf{|$d$|} \\
    \hline
    \hline
    \multirow{3}{*}{Age} & \multirow{3}{*}{\texttt{Hadoop}} & \texttt{REP} & 0.65 & 0.13 (negligible)\\
      & &  \texttt{Siamese Pair} & $<0.001$ & 0.98 (large) \\ 
      & & \texttt{SABD} & 0.013 & 0.67 (large) \\
    \hline
    \hline
    \multirow{6}{*}{State} & \multirow{3}{*}{\texttt{Hadoop}} & \texttt{REP} & 0.199 & 0.35 (medium)  \\
      & &  \texttt{Siamese Pair} &  0.520 & 0.18 (small)\\ 
      & & \texttt{SABD} & 0.796 & 0.08 (negligible)\\
    \cline{2-5}
      & \multirow{3}{*}{\texttt{Spark}} & \texttt{REP} & 0.143 & 0.4 (medium) \\
      & &  \texttt{Siamese Pair} & 0.058 & 0.51 (large) \\ 
      & & \texttt{SABD} & 0.043 & 0.54 (large) \\
    \hline
    \end{tabular}
    \label{tab:state-age-jira}
\end{table}

As mentioned in Section~\ref{sec:exp-setup} Experimental Setup, to investigate the age bias (old vs. recent data) and state bias (initial vs. latest state), we run experiments on the BRs from Bugzilla. However, it stays unknown whether age bias and state bias exist in another ITS. Here, we investigate the age and state bias on data from an alternative ITS other than Bugzilla.

\textbf{Age bias in an ITS other than Bugzilla}: In RQ1, the experimental results demonstrate that there is a significant difference between tools in research on the old data and recent data from Bugzilla. Besides Bugzilla, we also conducted experiments on the old data from Jira. 
In RQ1, for Jira, we selected \texttt{Hadoop} and \texttt{Spark}.
However, since the first issue from \texttt{Spark} project was created on December 18, 2013. 
There are not enough old BRs to investigate the age bias. 
Note that the same reason also applies to all projects selected in our work which uses GitHub ITS. 
Specifically, the first issue from \texttt{VScode} project was created on October 14, 2015 and the first issue from \texttt{Kibana} project was created on February 6, 2013. 
Therefore, we are only able to conduct experiments on the old data of \texttt{Hadoop}, which uses Jira as an ITS. 
Following what we did in the RQ1, we evaluate the three tools \texttt{REP}, \texttt{Siamese Pair}, and \texttt{SABD} on the \texttt{Hadoop} old dataset (which contains BRs submitted between 2012 and 2014). 
Table~\ref{tab:state-age-jira} shows the statistical test results of the performance of these tools in \texttt{Hadoop}'s old and recent data. According to the $p$-value, we can find that age bias is significant in most cases in \texttt{Hadoop}.

\textbf{State bias in an ITS other than Bugzilla}: In RQ1, the experimental results show that there is no significant difference between tools performed with the initial states and the latest states. Besides Bugzilla, we leverage the datasets shared by Montgomery et al.~\cite{montgomery2022alternative} and recover the states of issues in Jira (i.e., \texttt{Hadoop} and \texttt{Spark}) to the end of the submission day. We then perform the same experiments as RQ1 for state bias and report the results in Table~\ref{tab:state-age-jira}.  
As Table~\ref{tab:state-age-jira} shows, the state bias is also insignificant in most cases in \texttt{Hadoop} and \texttt{Spark}, which uses Jira ITS. 
Note that since the issue change history in GitHub can be deleted, the saved history may be incomplete. 
Thus, we did not investigate the state bias in GitHub.

\subsection{Performance Comparison of tools in research and practice on old data.}

\begin{figure*}
    \begin{minipage}[b]{0.5\textwidth}
    \centering
    \includegraphics[width=\textwidth]{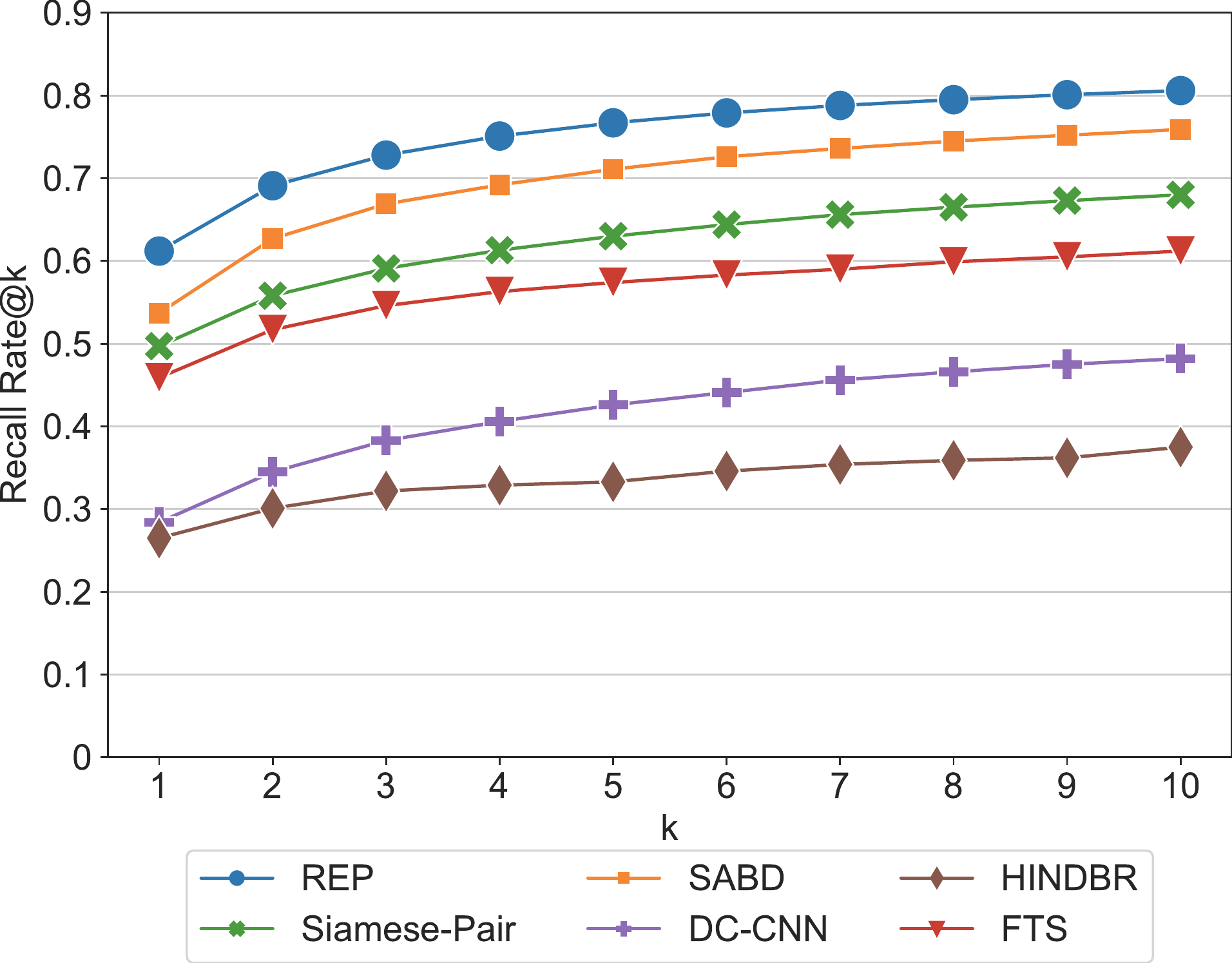}\\
    \subcaption{Eclipse Old}
    \end{minipage}%
    \begin{minipage}[b]{0.5\textwidth}
    \centering
    \includegraphics[width=\textwidth]{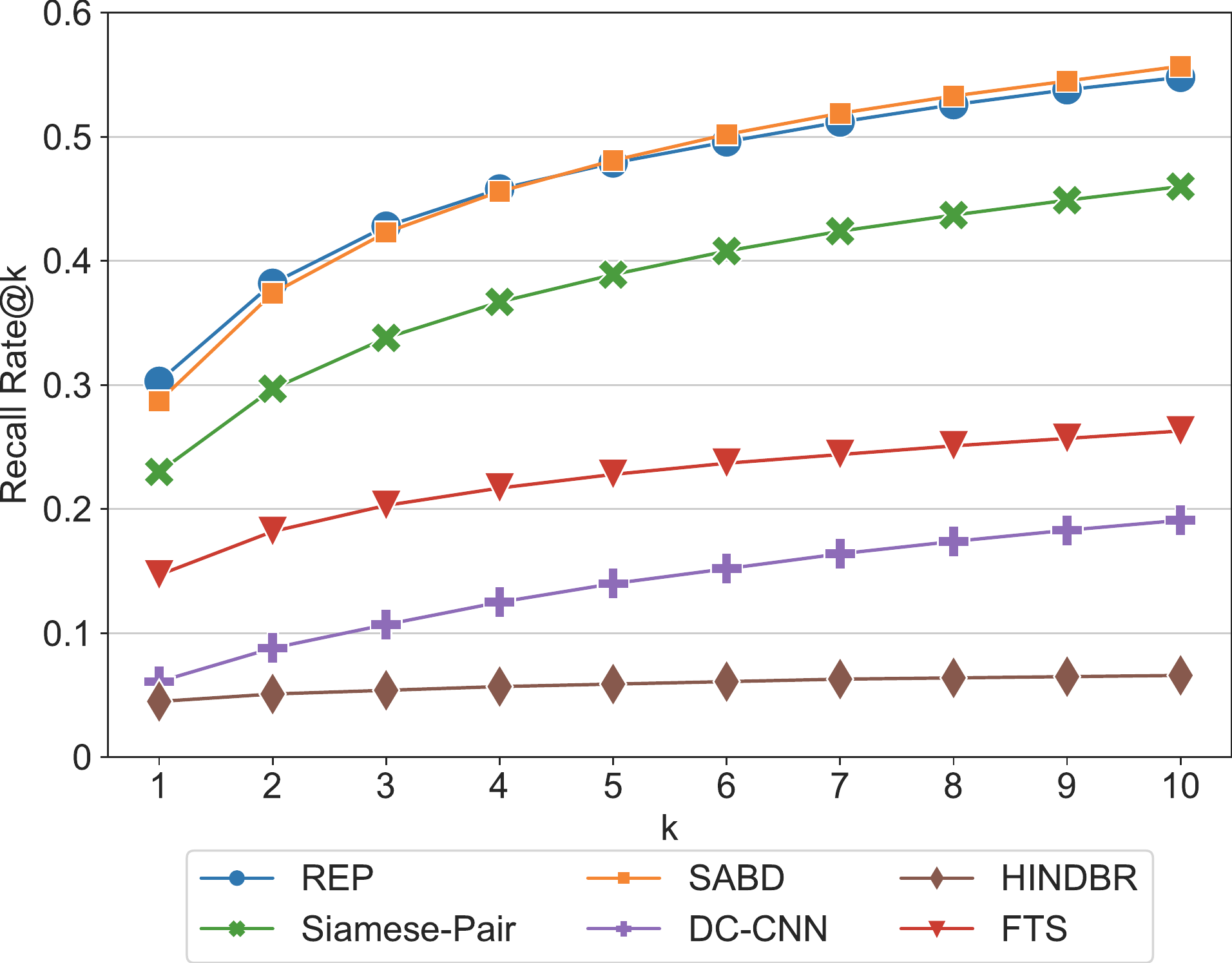}\\
    \subcaption{Mozilla Old}
    \end{minipage}%
\caption{Recall Rate@$k$ in the test data of Eclipse-Old, Mozilla-Old}
\label{fig:comp-old}
\end{figure*}

In RQ3, we demonstrate the performance comparison between tools in research and tools in practice. 
The experimental results show that \texttt{REP} and \texttt{SABD} are the best performers. 
\texttt{FTS} is better than \texttt{HINDBR} and \texttt{DC-CNN}. 
However, it remains unknown how different DBRD tools compare with each other in the data from old age. 
For the two tools in practice we evaluated, since \texttt{VSCodeBot} is not open-sourced, hence we can only investigate \texttt{FTS}. 
Figure~\ref{fig:comp-old} shows that \texttt{REP} and \texttt{SABD} are still the best performers. \texttt{Siamese Pair} comes in third. \texttt{FTS} is better than \texttt{HINDBR} and \texttt{DC-CNN} in both datasets. 
Even so, the performance of all the approaches dropped in \texttt{Mozilla}'s old dataset.
It suggests that when the size of a dataset becomes larger, the performances of DBRD approaches do not always improve. 

\subsection{Failure Analysis}
To understand what are the causes of DBRD approaches that failed to detect some duplicate BRs, we investigated the three best performers, i.e., \texttt{REP}, \texttt{SABD}, and \texttt{Siamese Pair}.
We selected the largest projects from each of the ITSs, i.e., \texttt{Mozilla}, \texttt{Hadoop}, and \texttt{VSCode}. We conducted the following steps to understand the causes of the DBRD failures:
\begin{enumerate}
    \item Firstly, we get the BRs that were not detected successfully in all five runs (out of 10 positions) by each approach in each dataset.
    \item Secondly, we sampled 50 BRs, which failed to recommend the three approaches on the three datasets.
\end{enumerate}

We identified 3 causes of failed duplicate detection and describe them as follows. 

(1) \textit{Limited or Incomplete description.} 
When the description is short, it does not provide enough context to understand the issue. Issue reporters attach screenshots or other supporting materials in the issue so that they neglected to write a detailed description. We also found that some descriptions contain too many URLs with only limited textual information. One such example is Bug 1668483~\footnote{\url{https://bugzilla.mozilla.org/show_bug.cgi?id=1668483}} from the \texttt{Mozilla} project. 
The description of this bug is full of long explicit URLs, which makes it hard for models to understand the real content in this issue.
Furthermore, we found that issue reporters may break the BR description into multiple parts. They can write a BR description into several comments, which were not considered by our work. One such example is Bug 1641043~\footnote{\url{https://bugzilla.mozilla.org/show_bug.cgi?id=1641043}} from the \texttt{Mozilla} dataset. 
The issue reporter actually described the issue in two consecutive parts, while only the first block is called ``description" and the second block is considered as ``comment". A complete version of the description may be helpful for DBRD approaches to detect duplicates.

(2) \textit{Inability of the current approaches to understand the different ingredients in BR descriptions.} Since the current DBRD approaches only treat the textual information as unstructured, they cannot extract useful information from the description. The useful information contained in the description may describe the failure, steps to reproduce, system information, etc. They are usually arranged in a structured way. Aside from natural language description, there can also be (1) code snippets, (2) logs, (4) backtraces inside the description. 
A more reasonable approach may be able to extract different types of information separately. One such example is the duplicate BR pair from the VScode dataset, i.e., issue \#105446~\footnote{\url{https://github.com/microsoft/vscode/issues/105446}} and issue \#110999~\footnote{\url{https://github.com/microsoft/vscode/issues/110999}}. Both issues contain the system information, steps to reproduce, and screenshots. However, since the information in the descriptions was not considered separately, it may be challenging for models to understand the failures. Other than textual information, a BR can also contain images or screen recordings in the description. However, the approaches evaluated in our work only consider the textual information in BR descriptions. Sometimes, the screenshot shows similar information. For instance, the duplicate issues \#108908~\footnote{\url{https://github.com/microsoft/vscode/issues/108908}} and \#107104~\footnote{\url{https://github.com/microsoft/vscode/issues/107104}} in VScode repository. Among these two issues, issue \#108908 describes the bug as ``overlap" and issue \#107104 describes the bug as ``loads them twice", which does not look duplicate for sure. However, based on the screenshots from both issues, we can find these two issues refer to the same bug. An approach that can handle the screenshots and screen recordings would be helpful in these cases.

(3) \textit{Different failures with the same underlying fault.} As indicated by Runeson et al.~\cite{runeson2007detection}, there are two types of duplicates: (1) they describe the same failure; (2) they describe two different failures with the same underlying fault. We also encountered difficult cases when both BR described the issue correctly, however, they described the two different failures while the underlying fault is the same. Since the current approaches are based on the similarity of BRs, it is challenging for them to detect the second type of duplicates. One such example is the duplicate BR pair from the Hadoop dataset, i.e., HBASE-24609~\footnote{\url{https://issues.apache.org/jira/browse/HBASE-24609}} and HBASE-24608~\footnote{\url{https://issues.apache.org/jira/browse/HBASE-24608}}. 
The two BRs describe two different objects, i.e., \texttt{MetaTableAccessor} and \texttt{CatalogAccessor}. Even for developers with some experience on Hadoop projects, it may not be possible to recognize that they are duplicates.

\subsection{Lessons Learned}
\vspace{0.2cm}\noindent{\bf Age bias and ITS bias should be considered for DBRD, and even for other tasks that involve BRs.}
We show that two kinds of bias (age and ITS) affect the performance of DBRD techniques.
These biases must be considered while designing and evaluating future DBRD techniques. Further, we believe that any task involving BRs, e.g., bug localization~\cite{pradel2020Scaffle, loyola2018bug, kim2021datasets}, bug severity prediction~\cite{arokiam2020automatically, liu2018predicting, baarah2019machine}, and bug triage~\cite{xuan2017automatic,su2021reducing} etc. should also provide due consideration for these biases. When evaluating an approach, it would be better to consider the diversity6 of the ITS. 

\vspace{0.2cm}\noindent{\bf Using \texttt{FTS} and \texttt{REP} as a baseline for evaluating DBRD approaches.} We observe that \texttt{FTS} although simple outperforms many other DBRD approaches for most projects (all except \texttt{Mozilla}). \texttt{REP}, although proposed a decade ago, is the overall best performer. Thus, we suggest future research include these simpler techniques as baselines. Future state-of-the-art approaches need to demonstrate superior performance over these simpler techniques.

\begin{table}[t]
\caption{Mann-Whitney-U with Cliff's Delta Effect Size $|d|$ on RQ1 with controlling the size of training/validation pairs} 
\label{tab:discuss_size} 
\centering
\small
    \begin{tabular}{|l|l|rr|l|r|r|}
    \hline
    \multirow{2}{*}{\textbf{Bias}}
    & \multirow{2}{*}{\textbf{Data}}
    & \multicolumn{2}{c|}{ \# \textbf{Sampled pairs}}
    & \multirow{2}{*}{\textbf{Approach}}
    & \multirow{2}{*}{\textbf{$p$-value}}
    & \multirow{2}{*}{\textbf{$|d|$}} \\
    & & \textbf{Training} & \textbf{Validation} & & & \\
    \hline
    \hline
    \multirow{6}{*}{Age} & \multirow{3}{*}{\texttt{Eclipse}} & \multirow{3}{*}{3,342} & \multirow{3}{*}{108} & \texttt{REP} & 0.002 & 0.78 (large) \\
    & & & & \texttt{Siamese Pair} & < 0.001 & 0.9 (large) \\
    & & & & \texttt{SABD} & 0.040 & 0.54 (large) \\
    \cline{2-7}
    & \multirow{3}{*}{\texttt{Mozilla}} & \multirow{3}{*}{68,396} & \multirow{3}{*}{2,418} & \texttt{REP} & 0.003 & 0.76 (large) \\
    & & & & \texttt{Siamese Pair} & 0.002 & 0.8 (large) \\
    & & & & \texttt{SABD} & 0.007 & 0.7 (large) \\
    \hline
    \hline
    \multirow{6}{*}{ITS} & \multirow{3}{*}{Jira} & \multirow{3}{*}{626} & \multirow{3}{*}{26} & \texttt{REP} & 0.047 & 0.37 (medium) \\
    & & & & \texttt{Siamese Pair} & $<0.001$ & 0.81 (large) \\
    & & & & \texttt{SABD} & $<0.001$ & 0.79 (large) \\
    \cline{2-7}
    & \multirow{3}{*}{GitHub} & \multirow{3}{*}{724} & \multirow{3}{*}{28} & \texttt{REP} & 0.029 & 0.41 (medium) \\
    & & & & \texttt{Siamese Pair} & $<0.001$ & 0.93 (large) \\
    & & & & \texttt{SABD} & 0.010 & 0.4775 (large) \\
    \hline
    \end{tabular}
\end{table}

\vspace{0.2cm}\noindent{\bf Choose your weapon - Projects with a medium to low volume of historical BRs may not benefit from deep learning-based tools.
}
The two best-performing tools are \texttt{REP} and \texttt{SABD}. \texttt{SABD} is deep learning-based, while \texttt{REP} is not.
Comparing the performance of both tools in Figure~\ref{fig:results}, we can notice that their performance is similar for projects with the largest number of BRs (\texttt{Mozilla} and \texttt{vscode}).
However, there is a clear big gap in performance for the other projects (although they still contain thousands of BRs as training data).
This suggests that the applicability of deep learning-based solutions may be limited to very large ITSs with tens of thousands of BRs submitted over a few year period (considering {\em age bias} and data drift phenomenon~\cite{salganicoff1997tolerating}). 
For most ITSs, non-deep learning-based approaches may outperform.
Note that, in our experiments, we did not use all the historical data for training since our findings in RQ1 show that there is a significant difference when applying a DBRD approach to old data and recent data. Besides, the old and recent data carry different characteristics, e.g., the number of BRs, so the predictions of the models trained in the past data may become less accurate in the recent data~\cite{vzliobaite2016overview}. In addition, when training with more data, the training process takes longer and is more computationally expensive.

As shown in Table~\ref{tab:our_dataset}, we identified that the number of BRs in different projects varies a lot (i.e., the number of training and validation pairs are different).
The size of training data might be a confounding factor.
To understand whether our findings still hold when we have the same number of training and validation pairs, we investigated the impact of the data size.
We adopted the pair generation strategy used by SABD~\cite{rodrigues2020soft}.
The positive pairs are all combinations of the BRs which belong to the same bucket.
On the other hand, the negative pairs are randomly generated by pairing a BR from one bucket with the BR from another bucket.
Since the number of positive pairs is fixed, we generated the same number of negative pairs.
For each bias and each dataset, we sampled the same number of training pairs and validation pairs.
For instance, when we work on age bias on the Eclipse dataset, as the old dataset has 8,668 BR pairs, while the recent dataset contains 3,342 BR pairs, we would sample the same number of 3,342 BR pairs from the old dataset (i.e., downsampling to the minority label).
The number of training and validation pairs are reported in Table~\ref{tab:discuss_size}.
We then conducted the same experiment for RQ1 with the sampled pairs as presented in Table~\ref{tab:discuss_size}.
The $p$-values regard each bias are all cases.
The detailed results can be found in our replication package.~\footnote{\url{https://github.com/soarsmu/TOSEM-DBRD}}
In the table, we find that the main message of this paper is still valid even if we work on the same number of training and validation pairs.
Please note that the experimental data for state bias have no difference in terms of size (i.e., the numbers of BRs for before and after state are the same), so we excluded it in this additional analysis.

\vspace{0.2cm}\noindent{\bf Future research approaches should compare with industry tools.}
Researchers have largely ignored the comparison of DBRD techniques with industry tools. We conducted experiments on both \texttt{FTS} and \texttt{vscodebot}. Our experiments showed that \texttt{FTS} and \texttt{VSCodeBot} can outperform many research tools. While we have highlighted the need for evaluation with industry tools in the context of DBRD, we believe, our suggestion is valid even for other software engineering tasks too. Researchers should investigate if some alternative tools are used in practice to solve the same/similar pain points and compare the performance of research tools with those ``defacto'' tools.

\begin{table*}[t]
\caption{Averaged seconds of per prediction used by different approaches in the test data of \texttt{Eclipse}, \texttt{Mozilla}, \texttt{Hadoop}, \texttt{Spark}, \texttt{Kibana}, and \texttt{VSCode}}
\label{tab:time} 
\centering
\small
\begin{tabular}{|l|l|l|l|l|l|l|l|}
    \hline
    \textbf{ITS} & \textbf{Project} & 
    \textbf{\texttt{REP}}  & \textbf{\texttt{Siamese Pair}} & \textbf{\texttt{SABD}} & \textbf{\texttt{DC-CNN}} & \textbf{\texttt{HINDBR}} & \textbf{\texttt{FTS}}\\
    \hline
    \hline
    \multirow{2}{*}{Bugzilla} 
    & \texttt{Eclipse} & 0.07 & 0.15 & 1.61 & 2.49 & 0.76 & 0.20 \\
    \cline{2-8}
    & \texttt{Mozilla} & 0.25 & 0.89 & 12.63 & 16.64 & 5.80 & 1.74\\
    \hline
    \multirow{2}{*}{Jira} 
    & \texttt{Hadoop} & 0.07 & 0.07 & 0.80 & 1.19 & 0.30 & 0.10\\
    \cline{2-8}
    & \texttt{Spark} & 0.13 & 0.05 & 0.69 & 1.02 & 0.28 & 0.08 \\
    \hline
    \multirow{2}{*}{GitHub} 
    & \texttt{Kibana} & 0.07 & 0.18 & 1.40 & 2.10 & 0.67 & 0.19 \\
    \cline{2-8}
    & \texttt{VSCode} & 0.08 & 0.70 & 3.53 & 5.56 & 1.92 & 0.40\\
    \hline
\end{tabular}
\end{table*}

\vspace{0.2cm}\noindent{\bf Efficiency matters for the pre-submission DBRD scenario.}
In the post-submission scenario, the DBRD technique has the liberty of time to predict duplicates, but it is not the case for the pre-submission scenario.
The DBRD response time varies depending on the number of BRs in the ITS.
JIT duplicate recommendation used by Bugzilla, i.e., \texttt{FTS}, works faster than most research tools as they only query the summary field of existing BRs.
In usability engineering, a response time of over 1 second is considered to interrupt a user’s flow of thought~\cite{nielsen1994usability}.
Given that users can perceive a delay difference of 100 milliseconds~\cite{brutlag2008user,nielsen1994usability}, some DBRD approaches, which take over 10 seconds to predict potential duplicates, do not seem to meet the requirements. 
We report the seconds per prediction spent by each approach.
The experiments were run on a machine with Intel(R) Xeon(R) Gold 5218R CPU @ 2.10GHz (Mem: 252G) with 4 GeForce RTX 2080Ti (11G).
Only one GPU was utilized when running a single deep learning model.
From Table~\ref{tab:time}, we can find that the time needed for each approach to make a prediction differs in different projects. 
Generally speaking, for tools in research, \texttt{REP}, and \texttt{Siamese Pair} are faster than the rest approaches, while in the largest project \texttt{Mozilla}, the run time difference is more pronounced. 
\texttt{REP} is $3.56\times$ faster than \texttt{Siamese Pair}, $50.52\times$ faster than \texttt{SABD}, and $66.56\times$ faster than \texttt{DC-CNN}. 
For the tools in practice, since \texttt{VSCodeBot} is not open-sourced, we cannot measure its run time on the pre-submission scenario. 
For \texttt{FTS}, we can find that it is faster than most research tools.

We suggest two potential topics for future DBRD research: (1) investigating the acceptable delay for the pre-submission DBRD scenario and (2) optimizing DBRD response time. 
To reduce the prediction time, future research can also consider reducing the search space. For example, instead of including all the BRs submitted within the one-year time window as candidates, further approaches can reduce the candidates first: applying a time-efficient technique, such as BM25, to filter out the BRs with a low chance of being duplicated. 
After that, expensive deep learning-based models can be used.

\textbf{Comments should be considered.} 
Based on our failure analysis, we found that comments in the BRs may be helpful. Especially, when the issue reporter separates the description of a BR into several parts. In the post-submission scenario, leveraging comments can provide additional information for DBRD tools to represent a BR.

\textbf{Different ingredients in description should be handled separately.} 
Although in Bugzilla and Jira, there are dedicated fields for categorical information, we found that the description can be arranged in a structured way. It can have steps to reproduce, expected behaviour, and observed behaviour etc. Issue reporters usually include images or videos inside the description. An approach that can understand different contents from the description would be beneficial. For GitHub, where the information such as system information, extensions used, and steps to reproduce, are usually included in the description, an approach that can extract all the useful information from the description would be more effective.

\textbf{Other resources in the project can be considered to further improve DBRD accuracy.} 
The current DBRD approaches are designed for detecting BRs with similar contents. If future approaches want to tackle the duplicates which have different failures, we suggest they consider other resources in the project, such as code base, to understand the relationship between different failures with the same root cause.

%% file: sec/7_threats.tex
\section{Threats to Validity}
\label{sec:threats}

\subsection{Threats to Internal Validity}
The threats to internal validity mainly relate to the correctness of our implementation.
In order to mitigate these threats, we use the replication packages provided by the authors \cite{rodrigues2020soft, he2020duplicate, xiao2020hindbr, sun2011towards}. 
Whenever the provided implementations are incomplete or unclear, we confirmed with the authors and perform the necessary steps.
For tools in research, we have managed to replicate all the prior works on their own datasets and obtain similar results as those that are reported in the original papers. 
For tools in practice, we implemented the \texttt{FTS} based on the source code provided by \texttt{Mozilla}~\footnote{\url{https://github.com/bugzilla/bugzilla/blob/230d73a11989a46b0a0d3f271a1c4a260f371bd7/Bugzilla/Bug.pm}}.
However, since we focus on the effectiveness of duplicate detection, we may not include all the optimizations that industry tools may apply to improve the run time efficiency.

Another threat relates to the quality of labels in our benchmark.
Our benchmark extracts duplicate BR relations from three different ITSs, i.e., Bugzilla, Jira, and GitHub.
For Bugzilla and Jira, BRs come with a field that allows developers to mark duplicate BRs as such explicitly.
Like many prior works~\cite{xie2018detecting,lazar2014generating,xiao2020hindbr}, our work regards the explicit labels that developers have recorded in BRs as ground truth.
Considering the relation of duplicate BRs is usually recorded by the developers who are familiar with the project, we believe the chance that these BRs were linked or unlinked incorrectly is low.
Different from Bugzilla and Jira, GitHub does not have a well-structured field to maintain duplicate references, so we implemented a regular expression to extract the duplicate issues based on GitHub's marking duplicates guideline.~\footnote{\url{https://docs.github.com/en/issues/tracking-your-work-with-issues/marking-issues-or-pull-requests-as-a-duplicate}}
However, the regular expression may introduce false positives (i.e., a comment seems like a duplicate issue reference, but actually not a duplicate).
To mitigate the threat, we sampled 384 issues from each \texttt{GitHub} project (i.e., \texttt{Kibana} and \texttt{VSCode}), and two of the authors manually investigated the correctness of duplicate labels.
We conclude that our regular expression is trustworthy as it showed 99.2\% and 94.5\% accuracy in the manual inspection for \texttt{Kibana} and \texttt{VSCode}, respectively.
Again, the reason that we did not check the false negatives for GitHub data is that almost in all the ITSs, it is unlikely to have complete duplicate BR datasets, as developers may forget to mark the duplicate BRs.
According to these manual investigation results, the number of mislabels on the duplicate BRs is low and they are not biased toward a specific baseline technique.
In other words, it is less likely to affect our ability to perform a fair comparison.
Thus, we believe this threat is minimal.
Another threat exists in evaluating deep learning models.
Due to the nature of deep learning based approaches, the results can vary among different executions.
We also acknowledge that the hyper-parameters used in our work may not be optimal. 
Their performance could possibly be better if one devotes more time to tuning the parameters.

Additionally, the tools used in practice may be optimized and updated timely, but the tools in research are not. 
This can be another threat to the results of our experiments, especially for RQ3.
To examine this potential threat, we have checked the commit history of the \texttt{FTS}.~\footnote{\url{https://github.com/bugzilla/bugzilla/commits/5.2/Bugzilla/Bug.pm}} 
We found that the latest commit related to the duplicate suggestion function was made on February 11, 2014. 
It shows that the FTS method is not updated timely. 
However, since the \texttt{VSCodeBot} is not open-sourced, we acknowledge that it is possible that the underlying model or algorithm used by \texttt{VSCodeBot} is updated timely. 
Even so, based on the experimental results in Figure~\ref{fig:rq3_vscodebot}, \texttt{VSCodeBot} did not perform better than \texttt{SABD} and \texttt{REP}.
Thus, we believe the effect of timely updates and optimization on the experimental results in RQ3 is minimal.

\subsection{Threats to External Validity}
The threats to external validity mainly relate to the generalizability of our findings.
In order to reduce these threats, we take into consideration three popular ITSs, while prior works generally only evaluate the proposed approaches using BRs from one ITS, i.e., Bugzilla. For each ITS, we also choose two large and well-maintained projects/repositories. 

Another potential threat to external validity is that our work only considers two non-deep learning-based approaches, and the finding may not be generalizable to other non-deep learning-based approaches. 
We found that most non-deep learning-based approaches were developed around a decade ago. Particularly, in the last 5 years, only very few non-deep learning-based techniques have been proposed and considered as baselines in the literature. Among the non-deep learning-based approaches and baselines, only \texttt{REP} is open-sourced.

\subsection{Threats to Construct Validity} 
The threats to construct validity mainly lie in the evaluation metrics we use in our work. 
To diminish these threats, we follow prior DBRD works~\cite{sun2011towards, rodrigues2020soft,zhou2012learning,kaushik2012comparative} and use the RR@$k$.

%% file: sec/8_relatedwork.tex
\section{Related Work}
\label{sec:related}

\subsection{DBRD Techniques and Practitioners' Perception}
Many studies have developed DBRD techniques in the past decade. The state-of-the-art approaches and popular ones have been described in Section~\ref{sec:dbrdinresearch}. 
Here, we introduce three classic works and a recent study assessing practitioners' perceptions about DBRD. 

One of the pioneer studies in DBRD is by Runeson et al.~\cite{runeson2007detection}.
They extracted textual fields in a BR (summary and description), and converted a BR into a vector of weights following the standard Vector Space Model (VSM)~\cite{schutze2008introduction}. 
Duplicates are then identified by comparing the vector representation of an incoming BR to those of existing ones wrt. three well-known similarity metrics: Cosine, Jaccard, and Dice~\cite{schutze2008introduction}. 
Wang et al.~\cite{wang2008approach} extended Runeson et al.'s work by considering both natural language text and execution information (e.g., stack traces). 
They have shown that the consideration of execution information is beneficial but not many BRs contain such execution information. 
Sun et al.~\cite{sun2010discriminative} trained a discriminative model via Support Vector Machine (SVM) to classify whether two BRs are duplicates of one other with a probability. Based on this probability score, they retrieve and rank candidate duplicate BRs. 

Zou et al.~\cite{zou2018practitioners} surveyed 327 practitioners from diverse backgrounds to investigate practitioners' perceptions of DBRD and other automated techniques supporting ITS. They find that DBRD is among the top-3 techniques deemed to be the most valued and find that respondents appreciate these techniques as they can save developers' time, save reporters' time, provide hints for bug fixing, etc.

\subsection{Evaluation of DBRD Techniques}
There are several studies on evaluating DBRD techniques~\cite{rakha2018revisiting,tu2018careful}. 
Rakha et al.~\cite{rakha2018revisiting} studied the differences between duplicate BRs before and after the introduction of JIT duplicate BR recommendation (JIT feature) in Bugzilla (in 2011). They found that duplicate BRs after 2011 (2012--2015) are less textually similar, have a greater identification delay, and require more discussion to be retrieved as BRs than duplicates before 2011. Their study also demonstrates that when evaluating the data after 2011, the experimental results of prior research would vary. These findings motivated us to experiment on BRs submitted after 2011. Based on their findings, we build the old and recent data both after 2011. Our work differs from their work as we investigate the two 3-year time window data after the JIT feature. We investigate not the impact of JIT DBRD feature introduction in Bugzilla, but rather {\em age bias}.

Tu et al.~\cite{tu2018careful} also highlighted the fact that the BR attributes, i.e., field values, change over time. They raise a concern that several DBRD techniques use data from the future while training their models. 
Their study can be seen dealing with our \textit{state} bias. 
They just investigate \texttt{REP}~\cite{sun2011towards}'s accuracy from BRs on Bugzilla ITS of Mozilla and Eclipse. 
In this work, we investigated diverse factors that could affect the performance of DBRD techniques, and not only the \textit{state} bias. 
Furthermore, we run a statistical test on the accuracy difference between using the initial state and the latest state, and show that state bias does not have a statistically significant impact.

%% file: sec/9_conclusion.tex
\section{Conclusion and Future Work}
\label{sec:conclusion}
In this work, we evaluated DBRD techniques both in research and practice. 
We analyzed the factors affecting DBRD performance. 
We showed that on recent data, DBRD approaches demonstrated significantly different performance when compared to their performance on old data.
Clearly, a DBRD technique that works well on data from many years ago but no longer so on recent data, should not be of much use to developers today.
Therefore, future research should use a recent data benchmark for evaluation.
We investigated three ITSs and two projects from each of them.
We propose that future research should consider GitHub for DBRD since the existing approaches do not perform as well on GitHub as they do on Bugzilla and Jira.
Taking a step ahead, we also compared the industry tools like Bugzilla's \texttt{FTS} and \texttt{VSCodeBot}.
We observe that although some tools proposed in research work perform better, \texttt{FTS} can serve as a strong baseline for comparison.
Furthermore, a DBRD technique, \texttt{REP}, proposed in 2011, outperforms advanced deep learning-based techniques that are recently proposed, and should also be used as a strong baseline.

Our work opens up exciting opportunities in DBRD.
We would like to broaden the ITSs and work with other ITSs such as the Android Issue Tracker.
As future work, we also plan to dig into the rich information available in GitHub issues, e.g., screenshots.
Leveraging more information other than pure text should help boost the accuracy of DBRD techniques.
Even though several DBRD techniques are available, they make no guarantees on response time (prediction time).
Developing a DBRD approach that works in real-time is another direction we wish to explore as future work. 